\documentclass[10pt]{article}
\usepackage{a4wide}
\usepackage{cite}
\usepackage{amsmath}
\usepackage{amsfonts}
\usepackage{graphicx}

\newcommand{\dd}{{\rm{d}}}        
\newcommand{\sign}{{\rm sign}}    
\newcommand{\Q}{{\mathcal{Q}}}    
\newcommand{\T}{{\mathcal{T}}}    
\newcommand{\HH}{{\mathcal{H}}}   
\newcommand{\PP}{{\mathcal{P}}}    
\newcommand{\QQ}{{\mathcal{Q}}}   
\newcommand{\scri}{{\mathcal{I}}} 

\newcommand{\sdfrac}[2]{\mbox{\small$\displaystyle\frac{#1}{#2}$}}

\def \bF {\mathbf{F}}
\def \bA {\mathbf{A}}

\newcommand{\rovno}{\!\!\!\!& = &\!\!\!\!}
\newcommand{\equi}{\!\!\!\!& \equiv &\!\!\!\!}

\begin{document}

\title{New improved form of black holes of type D}

\author{
Ji\v{r}\'{\i} Podolsk\'y
and
Adam Vr\'atn\'y\thanks{
{\tt podolsky@mbox.troja.mff.cuni.cz}
and
{\tt vratny.adam@seznam.cz}
}
\\ \ \\ \ \\
Institute of Theoretical Physics, Charles University, \\
V~Hole\v{s}ovi\v{c}k\'ach 2, 18000 Prague 8, Czech Republic.
}

\maketitle

\begin{abstract}
We derive a new metric form of the complete family of black hole spacetimes (without a cosmological constant) presented by Pleba\'nski and Demia\'nski in 1976. It further improves the convenient representation of this large family of exact black holes found in 2005 by Griffiths and Podolsk\'y. The main advantage of the new metric is that the key functions are considerably simplified, fully explicit, and factorized. All four horizons are thus clearly identified, and degenerate cases with extreme horizons can easily be discussed. Moreover, the new metric depends only on six parameters with direct geometrical and physical meaning, namely $m, a, l, \alpha, e, g$ which characterize mass, Kerr-like rotation, NUT parameter, acceleration, electric and magnetic charges of the black hole, respectively. This general metric reduces directly to the familiar forms of either (possibly accelerating) Kerr--Newman, charged Taub--NUT solution, or (possibly rotating and charged) $C$-metric by simply setting the corresponding parameters to zero, without the need of any further transformations. In addition, it shows that the Pleba\'nski--Demia\'nski family does not involve accelerating black holes with just the NUT parameter, which were discovered by Chng, Mann and Stelea in 2006. It also enables us to investigate various physical properties, such as the character of singularities, horizons, ergoregions, global conformal structure including the Penrose diagrams, cosmic strings causing the acceleration of the black holes, their rotation, pathological regions with closed timelike curves, or explicit thermodynamic properties. It thus seems that our new metric is a useful representation of this important family of black hole spacetimes of algebraic type~D in the asymptotically flat settings.
\end{abstract}

\vfil\noindent
PACS class:  04.20.Jb, 04.70.Bw, 04.40.Nr, 04.70.Dy



\bigskip\noindent
Keywords: black holes, exact spacetimes, accelerating and rotating sources, NUT charge, type D solutions, Pleba\'nski--Demia\'nski class
\vfil
\eject

\section{Introduction}
\label{intro}

In this contribution, we derive and analyze a new coordinate representation of the Pleba\'nski--Demia\'nski spacetimes \cite{PlebanskiDemianski:1976} describing a large class of black holes (identified also by Debever \cite{Debever:1971}). It contains, as special cases, all the well-known simpler black holes, namely the Schwarzschild (1915), Reissner--Nordstr\"{o}m (1916--1918), Kerr (1963), Taub--NUT (1963) or Kerr--Newman (1965) black holes, and also the $C$-metric (1918, 1962), physically interpreted by Kinnersley--Walker (1970) as uniformly accelerating pair of black holes, see e.g. \cite{Stephanietal:2003, GriffithsPodolsky:2009}. These accelerating black holes can also be charged, rotating, and can admit the NUT twist parameter.

The class of Pleba\'nski--Demia\'nski spacetimes, which includes all these famous black holes, is a family of exact solutions to Einstein--Maxwell equations of algebraic type~D with double-aligned non-null electromagnetic field (in the present paper we restrict ourselves only to the case of vanishing cosmological constant) --- see Chapter~16 of the monograph \cite{GriffithsPodolsky:2009} for the recent review and number of related references.

Our new form of the metric, which further improves the convenient representation of the class of Pleba\'nski--Demia\'nski black holes found by Griffiths and Podolsk\'y \cite{GriffithsPodolsky:2005,GriffithsPodolsky:2006,PodolskyGriffiths:2006}, reads
\begin{align}
\dd s^2 = \frac{1}{\Omega^2} &
  \left(-\frac{Q}{\rho^2}\left[\,\dd t- \left(a\sin^2\theta +4l\sin^2\!\tfrac{1}{2}\theta \right)\dd\varphi \right]^2
   + \frac{\rho^2}{Q}\,\dd r^2 \right. \nonumber\\
& \quad \left. + \,\frac{\rho^2}{P}\,\dd\theta^2
  + \frac{P}{\rho^2}\,\sin^2\theta\, \big[ a\,\dd t -\big(r^2+(a+l)^2\big)\,\dd\varphi \big]^2
 \right), \label{newmetricGP2005}
\end{align}
where
\begin{eqnarray}
\Omega    \rovno 1-\frac{\alpha\,a}{a^2+l^2}\, r\,(l+a \cos \theta) \,, \label{newOmega}\\
\rho^2    \rovno r^2+(l+a \cos \theta)^2 \,, \label{newrho}\\
P(\theta) \rovno \Big( 1-\frac{\alpha\,a}{a^2+l^2}\, r_{+} (l+a \cos \theta) \Big)
                  \Big( 1-\frac{\alpha\,a}{a^2+l^2}\, r_{-} (l+a \cos \theta) \Big) , \label{newP}\\
Q(r) \rovno \big(r-r_{+} \big) \big( r-r_{-} \big)
            \Big(1+\alpha\,a\,\frac{a-l}{a^2+l^2}\, r\Big)
            \Big(1-\alpha\,a\,\frac{a+l}{a^2+l^2}\, r\Big). \label{newQ}
\end{eqnarray}
The main roots of $Q(r)$, which identify the two black-hole horizons, are (independently of $\alpha$) located at
\begin{eqnarray}
r_{+} \equi  m +\sqrt{m^2 + l^2 - a^2 - e^2 - g^2}\,, \label{r+}\\
r_{-} \equi  m -\sqrt{m^2 + l^2 - a^2 - e^2 - g^2}\,, \label{r-}
\end{eqnarray}
with the (naturally positive) physical parameters
\begin{align}
m \quad.....\quad & \hbox{mass}               \nonumber \,,\\
a \quad.....\quad & \hbox{Kerr-like rotation} \nonumber \,,\\
l \quad.....\quad & \hbox{NUT parameter}      \nonumber \,,\\
e \quad.....\quad & \hbox{electric charge}    \nonumber \,,\\
g \quad.....\quad & \hbox{magnetic charge}    \nonumber \,,\\
\alpha \quad.....\quad & \hbox{acceleration}  \nonumber \,.
\end{align}

This is a further simplification of the previous Griffiths--Podolsk\'y form of the metric. The generic structure of the metric has remained basically the same (compare \eqref{newmetricGP2005} with Eq.~(16.18) in \cite{GriffithsPodolsky:2009}, renaming ${\PP\to P}$, ${\QQ\to Q}$ and ${\varrho \to \rho}$), but the new metric functions $P(\theta)$ and $Q(r)$ are now much more compact and explicit than previous ${\PP(\theta)}$ and ${\QQ(r)}$. They are nicely factorized, with $P$ determining the deficit angles corresponding to the cosmic strings along the axes ${\theta=0, \pi}$ of the black holes (causing the acceleration), while the roots of $Q$ clearly  determine the four horizons. Moreover, the ambiguous twist parameter $\omega$ has been removed by its most convenient fixing.

To see these improvements explicitly, let us recall the original Griffiths--Podolsk\'y form  \cite{GriffithsPodolsky:2005} of the metric functions, namely
\begin{align}
\Omega = 1 - \alpha\, \Big(\,\frac{l}{\omega}+\frac{a}{\omega}\cos\theta \Big)\, r\,, \qquad
\rho^2 = r^2 + (l + a \cos\theta)^2\,, \label{OmegaRho}
\end{align}
and
\begin{align}
\PP(\theta) &= 1-a_3\cos\theta-a_4\cos^2\theta\,,  \label{Poriginal}\\
\QQ (r) &= \Big[(\omega^2k+\tilde{e}^2+\tilde{g}^2)\Big(1+2\alpha \frac{l}{\omega}\,r\Big)
  -2\tilde{m}\,r +\frac{\omega^2k}{a^2-l^2}\,r^2\Big]
  \Big[1+\alpha\,\frac{a-l}{\omega}\,r\Big] \Big[1-\alpha\,\frac{a+l}{\omega}\,r\Big], \label{PfactorizedQ}
\end{align}
where the constants are
\begin{align}
a_3 &= 2\alpha \frac{a}{\omega}\,\tilde{m} -4\alpha^2 \frac{a\,l}{\omega^2}\,  (\omega^2k+\tilde{e}^2+\tilde{g}^2)\,,  \nonumber\\
a_4 &= -\alpha^2 \frac{a^2}{\omega^2}\,(\omega^2k+\tilde{e}^2+\tilde{g}^2)\,, \label{a34}
\end{align}
and $\omega^2 k$ is given by
\begin{equation}
  \frac{\omega^2 k}{a^2-l^2} =
  \frac{{\displaystyle 1 +2\alpha \frac{l}{\omega}\,\tilde{m} -3\alpha^2\frac{l^2}{\omega^2}\,(\tilde{e}^2+\tilde{g}^2)}}
  {{\displaystyle 1 + 3\alpha^2 \frac{l^2}{\omega^2}\,(a^2-l^2)}}\,,
  \label{k}
\end{equation}
which implies the expression
\begin{equation}
  \omega^2 k+\tilde{e}^2+\tilde{g}^2 =
  \frac{{\displaystyle (a^2-l^2 +\tilde{e}^2+\tilde{g}^2) +2\alpha \frac{l}{\omega}(a^2-l^2)\,\tilde{m} }}
  {{\displaystyle 1 + 3\alpha^2\frac{l^2}{\omega^2}\,(a^2-l^2)}}\,.
  \label{keg}
\end{equation}
Substituting  \eqref{a34}--\eqref{keg} into \eqref{Poriginal} and \eqref{PfactorizedQ} gives explicit but cumbersome expressions for the key metric functions $\PP(\theta)$ and $\QQ (r)$. This is now simplified in the new compact form of the metric \eqref{newmetricGP2005}--\eqref{newQ}.

\section{Derivation of the new metric}
\label{sec_derivation}

The first step in improving the form of the spacetime is to concentrate on the first factor of the metric function $\QQ (r)$ given by \eqref{PfactorizedQ}, which is quadratic in $r$. It can be rewritten as
\begin{align}
& \Big[(\omega^2k+\tilde{e}^2+\tilde{g}^2)\Big(1+2\alpha \frac{l}{\omega}\,r\Big)
  -2\tilde{m}\,r +\frac{\omega^2k}{a^2-l^2}\,r^2\Big]  \nonumber\\
& = \frac{\omega^2k}{a^2-l^2} \Big[ r^2 -2\tilde{m}\frac{a^2-l^2}{\omega^2k}\,r
  +\Big( 1 + 2\alpha \frac{l}{\omega}\,r \Big)
  \Big(a^2-l^2+\frac{a^2-l^2}{\omega^2k}(\tilde{e}^2+\tilde{g}^2)\Big)\Big]. \label{Qquadratic}
\end{align}
It can now be observed that this rather complicated expression nicely simplifies if we introduce a \emph{new set of the mass and charge parameters} $m, e,g$ in such a way that
\begin{eqnarray}
m   \equi \frac{a^2-l^2}{\omega^2 k}\,\tilde{m} - \alpha \frac{l}{\omega} (a^2-l^2  + e^2 + g^2)\,, \nonumber\\
e^2 \equi \frac{a^2-l^2}{\omega^2 k}\, \tilde{e}^2 \,, \label{PD-par-trans}\\
g^2 \equi \frac{a^2-l^2}{\omega^2 k}\, \tilde{g}^2 \,. \nonumber
\end{eqnarray}
Indeed, the factor \eqref{Qquadratic} then takes the explicit form
\begin{align}
\frac{\omega^2k}{a^2-l^2} \Big[\, r^2 - 2m\,r + (a^2-l^2+e^2+g^2)\Big]. \label{newQquadratic}
\end{align}
Provided ${m^2 + l^2 > a^2 + e^2 + g^2}$, it \emph{has two explicit roots} $r_+$ and $r_-$ given by \eqref{r+} and \eqref{r-}, respectively. The metric function \eqref{PfactorizedQ} can thus be factorized to
\begin{align}
\QQ (r) = S^{-1} \, \big(r-r_{+} \big) \big( r-r_{-} \big)
            \Big(1+\alpha\,\frac{a-l}{\omega}\,r\Big) \Big(1-\alpha\,\frac{a+l}{\omega}\,r\Big),
\label{factorizedQ}
\end{align}
where the \emph{constant}~$S$ is a shorthand for the inverse of \eqref{k}, namely
\begin{equation}
  S^{-1} \equiv \frac{\omega^2 k}{a^2-l^2} \,.
  \label{Sdef}
\end{equation}

Substitution from \eqref{PD-par-trans} into \eqref{k}, rewritten as
\begin{equation}
  \frac{\omega^2 k}{a^2-l^2}\Big[ 1 + 3\alpha^2\frac{l^2}{\omega^2}\,(a^2-l^2) \Big] =
  1 +2\alpha \frac{l}{\omega}\,\tilde{m} -3\alpha^2\frac{l^2}{\omega^2}\,(\tilde{e}^2+\tilde{g}^2)\,,
  \label{korig}
\end{equation}
yields the explicit expression for $S$ in terms of the new physical parameters
\begin{equation}
  S = 1 - 2\alpha \frac{l}{\omega}\, m  + \alpha^2 \frac{l^2}{\omega^2}\,(a^2-l^2+e^2+g^2)\,.
  \label{S}
\end{equation}
Notice that it can also be expressed in terms of the roots  $r_+$ and $r_-$  as
\begin{equation}
  S = 1 - \alpha \frac{l}{\omega}\,(r_+ +r_-)  + \alpha^2 \frac{l^2}{\omega^2}\,r_+ r_-
    = \Big(1 - \alpha \frac{l}{\omega}\,r_+ \Big) \Big(1 - \alpha \frac{l}{\omega}\,r_-\Big)\,.
  \label{Sroots}
\end{equation}

One may be worried about the change of the ``main physical parameters'' introduced by \eqref{PD-par-trans}. However, by inspecting the expressions \eqref{korig}, \eqref{S} it is immediately seen  that
\begin{align}
 \alpha \frac{l}{\omega}=0 \qquad& \hbox{implies} \qquad  S=\frac{a^2-l^2}{\omega^2 k} = 1\,, \nonumber\\
 & \hbox{and consequently} \quad  m=\tilde{m}\,, \ e=\tilde{e}\,,\ g=\tilde{g}\,.
  \label{k=1}
\end{align}
It means, that \emph{in all the subcases} ${\alpha=0}$ or ${l=0}$ (namely for Schwarzschild, Reissner--Nordstr\"{o}m, Kerr, Taub--NUT or Kerr--Newman black holes, and also for their accelerating generalizations with vanishing NUT parameter~$l$) the mass parameter $m$ and the charges $e,g$ actually \emph{remain the same}. And since there are no accelerating \emph{purely} NUT black holes in the Pleba\'nski--Demia\'nski class of type~D solutions, see \cite{PodolskyVratny:2020}, the difference between ${m, e, g}$ and ${\tilde{m}, \tilde{e}, \tilde{g}}$ occurs \emph{only if} ${\alpha\,a\, l \ne 0}$, cf.~\eqref{ChoiceOmega-al}. That is the most general case of \emph{accelerating} black holes \emph{with both the rotation $a$ and the NUT parameter~$l$}, whose geometric and physical properties have not yet been studied.

After factorizing the function $\QQ(r)$, as the second step we now turn to the metric function $\PP(\theta)$ determined by the constants $a_3$ and $a_4$. It is known that these two Pleba\'nski--Demia\'nski metric functions are related, and for vanishing cosmological constant they share the root structure. It can thus be expected that also the function $\PP(\theta)$ could be factorized by the suitable reparametrization \eqref{PD-par-trans}. This is indeed the case. Expressing \eqref{a34} in terms of the new parameters $m, e, g$ we get
\begin{align}
a_3 &= 2\alpha  \frac{a}{\omega}\,\frac{\omega^2 k}{a^2-l^2}\,
\Big[\,m - \alpha \frac{l}{\omega} (a^2-l^2  + e^2 + g^2)\Big]\,,  \nonumber\\
a_4 &= -\alpha^2 \frac{a^2}{\omega^2}\,\frac{\omega^2 k}{a^2-l^2}\,(a^2-l^2+e^2+g^2)\,. \label{a34new}
\end{align}
Using \eqref{Sdef}, \eqref{S} and substituting \eqref{a34new} into \eqref{Poriginal} we obtain
\begin{align}
\PP(\theta) &= S\frac{\omega^2 k}{a^2-l^2}-a_3\cos\theta-a_4\cos^2\theta  \nonumber\\
& = \frac{\omega^2 k}{a^2-l^2}\,\Big[\,1
   -2\alpha\,\frac{l+a\,\cos\theta}{\omega}\,m
   +\alpha^2 \frac{(l+a\,\cos\theta)^2}{\omega^2} (a^2-l^2  + e^2 + g^2) \Big] \nonumber\\
& = S^{-1} \Big(1 -\alpha\,r_+ \frac{l+a\,\cos\theta}{\omega}\Big)
           \Big(1 -\alpha\,r_- \frac{l+a\,\cos\theta}{\omega}\Big)\,.
\label{factorizedP}
\end{align}
The metric function $\PP(\theta)$ is thus also factorized when ${m^2 + l^2 > a^2 + e^2 + g^2}$, i.e., when the roots $r_+$ and $r_-$ exist.

To summarize, we have obtained the key expressions \eqref{factorizedQ} and \eqref{factorizedP}, which can be written as
\begin{align}
\QQ(r)      = S^{-1}\, Q (r)\,, \qquad
\PP(\theta) = S^{-1}\, P(\theta)\,, \label{QP relations}
\end{align}
where
\begin{align}
Q (r) &= \big(r-r_{+} \big) \big( r-r_{-} \big)
         \Big(1+\alpha\,\frac{a-l}{\omega}\,r\Big) \Big(1-\alpha\,\frac{a+l}{\omega}\,r\Big), \label{defnewQ}\\
P(\theta)&= \Big(1 -\alpha\,r_+ \frac{l+a\,\cos\theta}{\omega}\Big)
            \Big(1 -\alpha\,r_- \frac{l+a\,\cos\theta}{\omega}\Big)\,.
\label{defnewP}
\end{align}
Putting these into the original metric \cite{GriffithsPodolsky:2005,GriffithsPodolsky:2006,PodolskyGriffiths:2006}
(which has the same form as \eqref{newmetricGP2005} with $Q$, $P$ replaced by $\QQ$, $\PP$, respectively) we get
\begin{align}
\dd s^2 = \frac{S}{\Omega^2} &
  \left(-\frac{Q}{\rho^2}\,S^{-2}\left[\dd t- \left(a\sin^2\theta +4l\sin^2\!{\textstyle\frac{1}{2}\theta} \right)\dd\varphi \right]^2 + \frac{\rho^2}{Q}\,\dd r^2 \right. \nonumber\\
& \quad \left. + \frac{\rho^2}{P}\,\dd\theta^2
  + \frac{P}{\rho^2}\,\sin^2\theta\, S^{-2}\big[ a\dd t -\big(r^2+(a+l)^2\big)\,\dd\varphi \big]^2
 \right). \label{newmetric}
\end{align}

The third step in deriving the new metric is now based on an observation (first made in \cite{Vratny:2018}) that it is possible to \emph{rescale the coordinates~$t$ and~$\varphi$} by a constant scaling factor ${S \ne 0}$ (because their range has not yet been specified). In other words, we perform the transformation ${t \to S\,t}$ and ${\varphi \to S\,\varphi}$ which effectively removes the constants $S$ from the conformal metric ${\dd \hat{s}^2 \equiv S^{-1}\,\dd s^2}$. Moreover, a \emph{constant conformal factor} $S^{-1}$ does not change the geometry of the spacetime (recall also \eqref{k=1}, according to which ${S=1}$ whenever ${\alpha\,a\,\l=0}$). Therefore, the Pleba\'nski--Demia\'nski black-hole solutions can equivalently be represented by the metric ${\dd \hat{s}^2}$. Dropping the hat, we arrive at the metric \eqref{newmetricGP2005}.

In fact, this specific rescaling procedure removes the two coordinate singularities hidden in the expression \eqref{Sroots} for $S$ at ${\alpha \, l \, r_{\pm} = \omega}$, making our new metric form (\ref{newmetricGP2005})--(\ref{newQ}) somewhat richer.

To complete the derivation, it only remains to \emph{fix the remaining twist parameter $\omega$}. In the original Griffiths--Podolsk\'y form  of the metric \cite{GriffithsPodolsky:2005}, this was left as a free parameter which could be set to \emph{any} value (if at least one of the parameters $a$ or $l$ are non-zero, otherwise ${\omega\equiv0}$ --- see the discussion in \cite{GriffithsPodolsky:2005, GriffithsPodolsky:2006}) using the remaining coordinate freedom. This ambiguity is unfortunate since the metric explicitly contains non-unique $\omega$ coupled \emph{both} to the Kerr-like rotation~$a$ \emph{and} the NUT parameter~$l$. We can now improve this drawback. It was found in \cite{Vratny:2018}, and conveniently employed in \cite{MatejovPodolsky:2021}, that the most suitable gauge choice of the twist parameter is
\begin{align}
\omega \equiv \frac{a^2 + l^2}{a}\,, \label{ChoiceOmega}
\end{align}
so that
\begin{align}
\frac{a}{\omega} = \frac{a^2}{a^2 + l^2}\,, \qquad \frac{l}{\omega} = \frac{a\,l}{a^2 + l^2}\,. \label{ChoiceOmega-al}
\end{align}
Substituting this gauge into the expressions \eqref{OmegaRho}, \eqref{defnewP} and \eqref{defnewQ}, we obtain the explicit metric functions $\Omega$, $P$ and $Q$ presented in \eqref{newOmega}, \eqref{newP} and \eqref{newQ}, respectively. The new form of the metric \eqref{newmetricGP2005}--\eqref{newQ}, which nicely represents the large family of type D black holes, is thus completely derived.

\section{Main subclasses of type~D black holes}
\label{sec_subclasse}

When ${m^2 + l^2 > a^2 + e^2 + g^2}$, the new metric \eqref{newmetricGP2005}--\eqref{newQ} naturally generalizes the standard forms of the most important black hole solutions. These are now easily obtained by setting the corresponding physical parameters to zero.

\subsection{Kerr--Newman--NUT black holes (${\alpha=0}$\,: no acceleration)}

By setting the acceleration parameter $\alpha$ to zero, the functions \eqref{newOmega}, \eqref{newP} reduce to
${\Omega=1}$, ${P=1}$, so that the generic metric \eqref{newmetricGP2005} simplifies as
\begin{align}
\dd s^2 = &
  -\frac{Q}{\rho^2}\left[\,\dd t- \left(a\sin^2\theta +4l\sin^2\!{\textstyle\frac{1}{2}\theta} \right)\dd\varphi \right]^2 + \frac{\rho^2}{Q}\,\dd r^2  \nonumber\\
& \quad  + \,\rho^2\,\dd\theta^2
  + \frac{\sin^2\theta}{\rho^2}\, \big[\, a\,\dd t -\big(r^2+(a+l)^2\big)\,\dd\varphi \big]^2
 \,, \label{metric-alpha=0}
\end{align}
where
\begin{eqnarray}
Q(r)   \rovno \big(r-r_{+} \big) \big( r-r_{-} \big) \,, \label{Q-alpha=0}\\
\rho^2 \rovno r^2+(l+a \cos \theta)^2 \,. \label{rho-alpha=0}
\end{eqnarray}
The two roots of $Q(r)$ identify the two black-hole horizons located at
\begin{eqnarray}
r_{\pm} \equi  m \pm \sqrt{m^2 + l^2 - a^2 - e^2 - g^2}\,. \label{r+-}
\end{eqnarray}
Famous subcases are readily obtained, namely the black holes solution of Kerr--Newman (${l=0}$), charged Taub--NUT (${a=0}$), Kerr (${l=0}$, ${e=0=g}$), Reissner--Nordstr\"{o}m (${a=0}$, ${l=0}$), and Schwarzschild (${a=0}$, ${l=0}$, ${e=0=g}$).

\subsection{Accelerating Kerr--Newman black holes (${l=0}$\,: no NUT)}

Without the NUT parameter $l$, the new metric \eqref{newmetricGP2005} simplifies to
\begin{align}
\dd s^2 = \frac{1}{\Omega^2} &
  \left(-\frac{Q}{\rho^2}\left[\,\dd t - a\sin^2\theta\, \dd\varphi \right]^2 + \frac{\rho^2}{Q}\,\dd r^2 \right. \nonumber\\
& \quad \left. + \,\frac{\rho^2}{P}\,\dd\theta^2
  + \frac{P}{\rho^2}\,\sin^2\theta\, \big[ a\,\dd t - (r^2+a^2)\,\dd\varphi \big]^2
 \right), \label{metric-l=0}
\end{align}
where
\begin{eqnarray}
\Omega     \rovno 1-\alpha\, r\,\cos\theta \,, \label{Omega-l=0}\\
\rho^2     \rovno r^2+a^2\cos ^2\theta \,, \label{rho-l=0}\\
P(\theta) \rovno \big( 1-\alpha\,r_{+} \cos\theta \big)
                 \big( 1-\alpha\,r_{-} \cos\theta \big) , \label{P-l=0}\\
Q(r) \rovno \big(r-r_{+} \big) \big( r-r_{-} \big)
            \big(1+\alpha\,r\big) \big(1-\alpha\,r\big). \label{Q-l=0}
\end{eqnarray}
This is a compact factorized form of the class of accelerating, rotating, and charged black holes. The spacetime admits 4 horizons, namely two black hole horizons at ${r_{\pm}= m \pm \sqrt{m^2 - a^2 - e^2 - g^2}}$ and two acceleration horizons at ${\pm \alpha^{-1}}$. For vanishing charges (${e=0=g}$), it is equivalent to the rotating $C$-metric identified by Hong and Teo \cite{HongTeo:2005}. For vanishing acceleration (${\alpha=0}$), the standard form of Kerr--Newman solution in Boyer--Lindquist coordinates is recovered.

\subsection{Charged Taub--NUT black holes (${a=0}$\,: no rotation)}

By setting the Kerr-like rotation parameter $a$ to zero, the new metric \eqref{newmetricGP2005} considerably simplifies and \emph{becomes independent of the acceleration}~$\alpha$ (because the metric functions \eqref{newOmega}--\eqref{newQ} depend on $\alpha$ only via the product ${\alpha\,a}$). Indeed, ${\Omega=1}$, ${P=1}$, so that
\begin{align}
\dd s^2 = &
  -\frac{Q}{\rho^2} \big(\dd t - 4l\sin^2\!{\textstyle\frac{1}{2}\theta} \,\dd\varphi \big)^2 + \frac{\rho^2}{Q}\,\dd r^2
  + (r^2+l^2)\big(\dd\theta^2 + \sin^2\theta\, \dd\varphi^2 \big)\,, \label{metric-a=0}
\end{align}
where
\begin{eqnarray}
Q(r) \rovno \big(r-r_{+} \big) \big( r-r_{-} \big)  \,, \label{Q-a=0}\\
\rho^2  \rovno r^2+l^2 \,. \label{rho-a=0}
\end{eqnarray}
This explicitly demonstrates that \emph{there is no accelerating NUT black hole in the Pleba\'nski--Demia\'nski family} of spacetimes. This observation was made already in the original works \cite{GriffithsPodolsky:2005,GriffithsPodolsky:2006,PodolskyGriffiths:2006}, and recently clarified. It was proven in \cite{PodolskyVratny:2020} that the metric for accelerating (non-rotating) black holes with purely NUT parameter --- which was found in 2006 by  Chng, Mann and Stelea \cite{ChngMannStelea:2006} and analyzed in detail in \cite{PodolskyVratny:2020} --- is of algebraic type~I. Therefore, it \emph{cannot} be contained in the Pleba\'nski--Demia\'nski class which is of type~D.

The charged Taub--NUT spacetime \eqref{metric-a=0} is nonsingular (its curvature does not diverge at ${r=0}$), away from the axis ${\theta=\pi}$ (where the rotating cosmic string is located) it is asymptotically flat as ${r\to\pm\infty}$, and the interior of the black hole is located between the two horizons ${r_+>0}$ and ${r_->0}$, where ${r_{\pm}= m \pm \sqrt{m^2 +l^2 - e^2 - g^2}}$.

\subsection{Uncharged accelerating Kerr--NUT black holes (${e=0=g}$\,: vacuum)}

Another nice feature of our new metric \eqref{newmetricGP2005}--\eqref{newQ} is that it \emph{has the same form for vacuum spacetimes} without the electromagnetic field. Indeed, the electric and magnetic charges $e$ and $g$, which generate the electromagnetic field, \emph{enter only the expressions for}~$r_\pm$  introduced in \eqref{r+}, \eqref{r-}. In other words, $e$ and $g$ just change the positions of the two black hole horizons. In the vacuum case, these constant parameters simplify to
\begin{eqnarray}
r_{\pm} \equi  m \pm \sqrt{m^2 + l^2 - a^2}\,. \label{r+-vacuum}
\end{eqnarray}
The metric \eqref{newmetricGP2005}--\eqref{newQ} with \eqref{r+-vacuum} represents the full class of accelerating Kerr--NUT black holes. It reduces to accelerating Kerr black hole when ${l=0}$, and non-accelerating Kerr--NUT black hole when ${\alpha=0}$. For ${a=0}$ it simplifies \emph{directly} to the Taub--NUT black hole \eqref{metric-a=0} without acceleration.

\section{Extreme black holes and hyperextreme cases}
\label{sec_extreme}

The new form of the generic black hole \eqref{newmetricGP2005} --- and also all its subclasses --- naturally admits a special case with a \emph{degenerate} horizon, which is the situation when the \emph{two horizons coincide}, ${r_{+}=r_{-}}$.
In view of \eqref{r+}, \eqref{r-}, this occurs if and only if the \emph{extremality condition}
\begin{equation}
m^2 + l^2 =  a^2 + e^2 + g^2 \label{extremality condition}\\
\end{equation}
is satisfied, and in such a case the \emph{extremal horizon is located at}
\begin{equation}
r =  m\,. \label{r-extreme}
\end{equation}
Consequently, the metric functions take the form
\begin{eqnarray}
P(\theta) \rovno \Big( 1-\frac{\alpha\,a\,m}{a^2+l^2}\, (l+a \cos \theta) \Big)^2 , \label{extremeP}\\
Q(r) \rovno (r-m)^2\,
            \Big(1+\alpha\,a\,\frac{a-l}{a^2+l^2}\, r\Big)
            \Big(1-\alpha\,a\,\frac{a+l}{a^2+l^2}\, r\Big), \label{extremeQ}
\end{eqnarray}
while all the remaining expressions in the metric \eqref{newmetricGP2005} remain the same. Apart from the degenerate black hole horizon at ${r=m}$ with zero surface gravity (and thus zero temperature), there are two acceleration horizons.

This large family of \emph{extremal accelerating Kerr--Newman--NUT black holes} admits various natural subclasses which are easily obtained by setting the corresponding physical parameters $\alpha, l, a, e, g$ to zero. In particular, Kerr--Newman--NUT black holes \emph{without acceleration} (${\alpha=0}$) take the form
\begin{align}
\dd s^2 = &
  -\frac{Q}{\rho^2}\left[\,\dd t- \left(a\sin^2\theta +4l\sin^2\!{\textstyle\frac{1}{2}\theta} \right)\dd\varphi \right]^2 + \frac{\rho^2}{Q}\,\dd r^2  \nonumber\\
& \quad  + \,\rho^2\,\dd\theta^2
  + \frac{\sin^2\theta}{\rho^2}\, \big[\, a\,\dd t -\big(r^2+(a+l)^2\big)\,\dd\varphi \big]^2
 \,, \label{extreme-metric-alpha=0}
\end{align}
where
\begin{eqnarray}
\frac{Q}{\rho^2} = \frac{(r-m)^2}{r^2+(l+a \cos \theta)^2} \,. \label{extremeQrho-alpha=0}
\end{eqnarray}
The subcases are Kerr--Newman (${l=0}$), charged Taub--NUT (${a=0}$), Kerr (${l=0}$, ${e=0=g}$), and Reissner--Nordstr\"{o}m (${a=0}$, ${l=0}$) extremal black holes, satisfying the extremality condition \eqref{extremality condition}.

Interestingly, in our recent work \cite{MatejovPodolsky:2021} we proved the equivalence of degenerate horizons in this family \eqref{extreme-metric-alpha=0}, \eqref{extremeQrho-alpha=0} of type~D black holes to a complete class of extremal isolated horizons with axial symmetry.

Finally, if the physical parameters satisfy the relation
\begin{equation}
m^2 + l^2 <  a^2 + e^2 + g^2 \,, \label{hyperextremality condition}\\
\end{equation}
the \emph{black hole horizons are absent}. This case represents \emph{hyperextreme} spacetimes with very large rotation $a$ and/or charges $e, g$. The metric function $Q(r)$ does not admit the real roots $r_+, r_-$. Instead, it involves a non-factorizable quadratic term of the form \eqref{newQquadratic}. In such a case, the metric  \eqref{newmetricGP2005} remains valid, but its metric functions $P$ and $Q$ are
\begin{eqnarray}
P(\theta) \rovno 1
   -2\,\alpha\,a\,\frac{l+a\,\cos\theta}{a^2+l^2}\,m
   +\alpha^2a^2 \frac{(l+a\,\cos\theta)^2}{(a^2+l^2)^2} (a^2-l^2  + e^2 + g^2) \, , \label{hyperextremeP}\\
Q(r) \rovno \big(r^2 - 2m\,r + (a^2-l^2+e^2+g^2)\big)
            \Big(1+\alpha\,a\,\frac{a-l}{a^2+l^2}\, r\Big)
            \Big(1-\alpha\,a\,\frac{a+l}{a^2+l^2}\, r\Big). \label{hyperextremeQ}
\end{eqnarray}
This exact spacetime represents a \emph{naked singularity} of mass $m$ with rotation $a$, NUT parameter $l$, electromagnetic charges $e,g$, and acceleration $\alpha$ caused by the tension of rotating cosmic strings attached to it along the axes. There are only two acceleration horizons. For ${\alpha=0}$, the metric simplifies considerably to the form \eqref{extreme-metric-alpha=0} with
\begin{eqnarray}
\frac{Q}{\rho^2} = \frac{r^2 - 2m\,r + (a^2-l^2+e^2+g^2)}{r^2+(l+a \cos \theta)^2} \,. \label{alpha=0hyperextremeQ}
\end{eqnarray}

The new metric form \eqref{newmetricGP2005}--\eqref{newQ} thus nicely describes the complete family of black holes of type~D, as well as their extreme cases and hyperextreme spacetimes with naked singularities.

\section{Physical discussion of the new metric}
\label{sec_discussion}

To study the global structure of the spacetime and to analyze its physical properties, it is first necessary to determine the gravitational field, in particular the specific curvature of the geometry, and the electromagnetic field. These are encoded in the Newman--Penrose scalars --- the components of the Riemann and Maxwell tensors with respect to the null tetrad. Its most natural choice~is
\begin{eqnarray}
\textbf{k} \rovno \frac{1}{\sqrt{2}}\, \frac{\Omega}{\rho} \bigg[ \frac{1}{\sqrt{Q}}
 \Big(\big(r^2+(a+l)^2\big)\, \partial_t + a \, \partial_\varphi \Big) + \sqrt{Q} \, \partial_r \bigg] \,, \nonumber \\
\textbf{l} \rovno \frac{1}{\sqrt{2}}\, \frac{\Omega}{\rho} \bigg[ \frac{1}{\sqrt{Q}}
 \Big(\big(r^2+(a+l)^2\big)\, \partial_t + a \, \partial_\varphi \Big) - \sqrt{Q} \, \partial_r \bigg] \,,  \label{nullframe}\\
\textbf{m} \rovno \frac{1}{\sqrt{2}}\, \frac{\Omega}{\rho} \bigg[
 \frac{1}{\sqrt{P} \sin \theta} \Big( \partial_\varphi + \big(a\sin^2\theta +4l\sin^2\!\tfrac{1}{2}\theta \big)\, \partial_t \Big)
 + \mathrm{i} \, \sqrt{P} \, \partial_\theta \bigg] \,. \nonumber
\end{eqnarray}
A direct calculation reveals that the only nontrivial Newman--Penrose scalars corresponding to the Weyl and Ricci tensors are
\begin{eqnarray}
\Psi_2 \rovno \frac{\Omega^3}{\big[r+\mathrm{i}\,(l+a \cos \theta) \big]^3} \bigg[ -(m+\mathrm{i}\,l)\Big(1-\mathrm{i}\,\alpha\, a\,\frac{a^2-l^2}{a^2+l^2}\Big)  \nonumber\\
&& \hspace{25mm}
+\frac{(e^2+g^2)}{r-\mathrm{i}\,(l+a \cos \theta)}
\Big(1+\frac{\alpha\,a}{a^2 + l^2} \big[a\,r \cos \theta +\mathrm{i}\, l\, (l+a \cos \theta) \big]\Big)
 \bigg]\,,
 \label{Psi2}\\
\Phi_{11} \rovno \tfrac{1}{2}(e^2+g^2) \,\frac{\Omega^4}{\rho^4} \,,
\label{Phi11}
\end{eqnarray}
while the Ricci scalar vanishes (indeed, ${R=0}$ for electrovacuum solutions with ${\Lambda=0}$).
Recall also \eqref{newOmega}, \eqref{newrho}, i.e.,
\begin{equation}
\Omega  = 1-\frac{\alpha\,a}{a^2+l^2}\, r\,(l+a \cos \theta) \,, \qquad
\rho^2  = r^2+(l+a \cos \theta)^2 \,. \label{newrho-rep}\\
\end{equation}

The curvature for the main subclasses of type~D black holes, summarized in Sec.~\ref{sec_subclasse}, are now easily obtained by setting up the corresponding physical parameters to zero:
\vspace{2mm}

\noindent
$\bullet$ {\bf Kerr--Newman--NUT  (${\alpha=0}$\,: no acceleration)}
\begin{eqnarray}
\Psi_2 \rovno \frac{1}{\big[r+\mathrm{i}\,(l+a \cos \theta) \big]^3} \bigg[ -(m+\mathrm{i}\,l)
+\frac{e^2+g^2}{r-\mathrm{i}\,(l+a \cos \theta)}
 \bigg]\,,
\label{Psi2-alpha=0}
\end{eqnarray}

\noindent
$\bullet$ {\bf Accelerating Kerr--Newman  (${l=0}$\,: no NUT)}
\begin{eqnarray}
\Psi_2 \rovno \frac{(1-\alpha\,r\cos\theta )^3}{(r+\mathrm{i}\,a \cos \theta )^3} \bigg[ -m\,(1-\mathrm{i}\,\alpha\, a)  + (e^2+g^2) \frac{1+\alpha\,r \cos \theta}{r-\mathrm{i}\,a \cos \theta}
 \bigg]\,,
\label{Psi2-l=0}
\end{eqnarray}

\noindent
$\bullet$ {\bf Charged Taub--NUT  (${a=0}$\,: no rotation)}
\begin{eqnarray}
\Psi_2 \rovno - \frac{m+\mathrm{i}\,l }{(r+\mathrm{i}\,l)^3}
  + \frac{e^2+g^2}{(r^2+l^2)(r+\mathrm{i}\,l)^2}\,.
\label{Psi2-a=0}
\end{eqnarray}

Of course, these expressions further simplify if (some of) the remaining parameters are zero. In particular, the \emph{Kerr--Newman} black hole is recovered from  \eqref{Psi2-alpha=0} if ${l=0}$. The \emph{C-metric} (accelerating charged black holes without rotation) are obtained from  \eqref{Psi2-l=0} when ${a=0}$. The \emph{Reissner--Nordstr\"om} black hole follows from \eqref{Psi2-a=0} when ${l=0}$. The \emph{uncharged} (vacuum) black holes are obtained for ${e=0=g}$. Moreover, all these particular expressions for $\Psi_2$ agree with those presented in the corresponding chapters of the monograph \cite{GriffithsPodolsky:2009}.

It is also useful to calculate the \emph{spin coefficients} for the null tetrad \eqref{nullframe}. It turns out that
\begin{eqnarray}
\kappa  \rovno \nu = 0\,, \qquad \sigma = \lambda = 0\,, \nonumber\\
\varrho \rovno \mu = -\frac{\sqrt{Q}}{\sqrt{2}\, \rho^3} \,
\Big( 1+\mathrm{i} \, \frac{\alpha\, a}{a^2 + l^2}\,(l+a \cos \theta)^2 \Big) \big(r-\mathrm{i}\,(l+a \cos \theta) \big) \,, \label{spincoef} \\
\tau \rovno \pi = -\frac{a\, \sqrt{P} \, \sin \theta}{\sqrt{2}\, \rho^3} \,
\Big( 1-\mathrm{i} \, \frac{\alpha\, a}{a^2 + l^2}\, r^2 \Big) \big(r-\mathrm{i}\,(l+a \cos \theta) \big) \,. \nonumber
\end{eqnarray}
Also the coefficients  ${\alpha = \beta}$ and ${\epsilon = \gamma}$ are non-zero (we do not write them because they are not simple). Both double-degenerate principal null directions generated by $\textbf{k}$ and $\textbf{l}$ are thus geodetic and shear-free. However, they have \emph{expansion} and \emph{twist} given  by
${\varrho = \mu \equiv - (\Theta + \mathrm{i}\, \omega)}$, that is
\begin{eqnarray}
\Theta \rovno \frac{\sqrt{Q} }{\sqrt{2} \,\rho^3} \,\Big(r+\frac{\alpha\, a}{a^2+l^2}\,(l+a \cos \theta)^3\Big)\,,\\
\omega \rovno - \frac{\Omega \sqrt{Q} }{\sqrt{2} \,\rho^3} \, (l+a \cos \theta)\,.
\end{eqnarray}
It is now explicitly seen that these \emph{black-hole spacetimes of algebraic type~D are non-twisting} (for a general  $r, \theta$) \emph{if and only if ${a=0=l}$}. Moreover, \emph{on the horizons} identified by ${Q(r)=0}$, both the expansion and the twist \emph{vanish} (${\Theta=0=\omega}$).

For investigation of the curvature singularities and asymptotically flat regions, it is also useful to evaluate the \emph{Kretschmann scalar}
\begin{eqnarray}
\mathcal{K} \equi R_{abcd}\, R^{abcd} = 48 \, {\cal R}e\,(\Psi_2^2)\,,
\end{eqnarray}
for type~D spacetimes. Interestingly, it takes the factorized form
\begin{equation}
\mathcal{K} = 48 \, \frac{\Omega^6}{\rho^{12}}\, K_+\,K_- \,, \label{Kretschmann}
\end{equation}
where
\begin{eqnarray}
K_\pm \rovno m \Big( F_\pm \pm \alpha\,a\,\frac{a^2-l^2}{a^2+l^2}\, F_\mp \Big)
 \mp l \Big( F_\mp \mp \alpha\,a\,\frac{a^2-l^2+e^2+g^2}{a^2+l^2}\, F_\pm \Big) \nonumber\\
&& - (e^2+g^2)\Big(1+\frac{\alpha\,a}{a^2+l^2}\,r L\Big) \, T_\pm \,, \\
F_\pm \rovno \big(r \mp L \big) \big( r^2 \pm 4 r L + L^2 \big)\,,  \qquad
T_\pm = \big( r^2 \pm 2 r L - L^2 \big)\,,  \qquad
L = l+a \cos \theta\,.  \nonumber
\end{eqnarray}
These expressions characterize the gravitational field.

When $e, g$ are not zero, the black-hole spacetime also contains a specific \emph{electromagnetic field} represented by the Maxwell 2-form ${\bF = \tfrac{1}{2}F_{ab}\, \dd x^a \wedge \dd x^b =\dd\bA}$. Its 1-form potential ${\bA = A_a \dd x^a}$~is
\begin{equation}
 \bA =  - \frac{e\,r + g\,(l+a \cos \theta)}{r^2+(l+a \cos \theta)^2}\, \dd t
 + \frac{(e\,r + g\,l)\,(a \sin ^2 \theta + 4 l \sin ^2 \tfrac{1}{2}\theta)
+ g\,\big(r^2+(a+l)^2\big)\cos \theta}{r^2+(l+a \cos \theta)^2} \,\dd\varphi\,.
\label{vector-potential}
\end{equation}
Therefore, the non-zero components of ${F_{ab} = A_{b,a}-A_{a,b}}$ are
\begin{eqnarray}
F_{rt}      \rovno \rho^{-4}\,\Big[\, e\,\big( r^2-(l+a \cos \theta)^2 \big) + 2\,g\,r\,(l+a \cos \theta) \Big]\,, \nonumber \\
F_{\varphi \theta} \rovno \rho^{-4}\,\Big[\, g\,\big( r^2-(l+a \cos \theta)^2 \big) - 2\,e\,r\,(l+a \cos \theta) \Big] \big(r^2 + (a+l)^2\big)\sin \theta  \,, \label{Fab}\\
F_{\varphi r} \rovno \big( a \sin ^2 \theta + 4 l \sin ^2 \tfrac{1}{2}\theta \big)\, F_{rt} \,, \nonumber \\
F_{\theta t} \rovno \frac{a}{r^2 + (a+l)^2}\,F_{\varphi \theta} \,. \nonumber
\end{eqnarray}
The corresponding Newman--Penrose scalars are ${\Phi_0 \equiv F_{ab}\, k^a m^b =  0}$,
${\Phi_2 \equiv F_{ab}\, \bar{m}^a l^b =0 }$, and
\begin{equation}
\Phi_1 \equiv \tfrac{1}{2} F_{ab}( k^a l^b + \bar{m}^a m^b)
   = \frac{\frac{1}{2}(e+\mathrm{i}\,g) \,\Omega^2}{\big(r+\mathrm{i}\,(l+a \cos \theta) \big)^2} 
   \,.
   \label{Phi1}
\end{equation}
It follows that ${\Phi_{11} = 2\,\Phi_1 \bar\Phi_1}$, in fully agreement with \eqref{Phi11}.

\subsection{Position of the horizons}
\label{subsec:horizon}

The new metric form (\ref{newmetricGP2005}) is very convenient for investigation of horizons. Clearly, the ``radial'' coordinate~$r$ is spatial in the regions where ${Q(r)>0}$, while it is a temporal coordinate where ${Q(r)<0}$. These regions are separated by horizons localized at ${Q(r)=0}$. In the case when ${m^2 + l^2 > a^2 + e^2 + g^2}$, the metric function~$Q$ is given by (\ref{newQ}),
\begin{eqnarray}
Q(r) \rovno \big(r-r_{+} \big) \big( r-r_{-} \big)
            \Big(1+\alpha\,a\,\frac{a-l}{a^2+l^2}\, r\Big)
            \Big(1-\alpha\,a\,\frac{a+l}{a^2+l^2}\, r\Big). \label{newQrep}
\end{eqnarray}
It is a \emph{quartic expression explicitly factorized into four real roots}, so that there are \emph{four horizons}, namely
\begin{eqnarray}
\HH_b^+ \quad \hbox{at} \quad  r_b^+ \equi r_{+}
   = m +\sqrt{m^2 + l^2 - a^2 - e^2 - g^2}\,, \label{r+rep}\\[1mm]
\HH_b^- \quad \hbox{at} \quad  r_b^- \equi r_{-}
   = m -\sqrt{m^2 + l^2 - a^2 - e^2 - g^2}\,, \label{r-rep}\\[2mm]
\HH_a^+ \quad \hbox{at} \quad  r_a^+ \equi +\frac{1}{\alpha}\,\frac{a^2+l^2}{a^2+a\,l}\,, \label{ra+}\\
\HH_a^- \quad \hbox{at} \quad  r_a^- \equi -\frac{1}{\alpha}\,\frac{a^2+l^2}{a^2-a\,l}\,, \label{ra-}
\end{eqnarray}
see the definitions of $r_\pm$ introduced in (\ref{r+}), (\ref{r-}). It is clear that ${r_{+}>0}$ \emph{for an arbitrary choice} of the physical parameters (assuming ${m>0}$), but ${r_{-}}$ \emph{can take any sign}. In particular,
\begin{eqnarray}
r_{-}>0    &\quad\Leftrightarrow \qquad    l^2 < a^2 + e^2 + g^2\,, \label{r-rep>0}\\
r_{-}<0    &\quad\Leftrightarrow \qquad    l^2 > a^2 + e^2 + g^2\,, \label{r-rep<0}\\
r_{-}=0    &\quad\Leftrightarrow \qquad    l^2 = a^2 + e^2 + g^2\,. \label{r-rep=0}
\end{eqnarray}

The horizons ${\HH_b^\pm}$ at ${r_b^\pm}$ are \emph{two black-hole horizons}. Interestingly, in our new metric form these are \emph{independent of the acceleration parameter~$\alpha$}. In fact, they are located \emph{at the same values} ${r_{+}, r_{-}}$ as the two horizons in the class of standard (non-accelerating) \emph{Kerr--Newman--NUT black holes} given by ${\alpha=0}$, see \cite{GriffithsPodolsky:2009}.

The horizons ${\HH_a^\pm}$ at ${r_a^\pm}$ are \emph{two acceleration horizons}. Their presence is the consequence of the fact that the black holes accelerate whenever the parameter ${\alpha}$ is non-zero. It is interesting that their location is now \emph{independent of mass~$m$ and charges~$e,g$} of the black holes. The values of ${r_a^\pm}$ depend only on the acceleration $\alpha$ and the specific combination of the twist parameters ${a,l}$. Moreover, when ${l=0}$ these are simply given just by the acceleration parameter as ${r_a^\pm = \pm \alpha^{-1}}$. They retain \emph{the same values} as in the \emph{$C$-metric} \cite{GriffithsPodolsky:2009} even if it is generalized to include the charges and rotation, that is for \emph{accelerating Kerr--Newman black holes}.

Of course, there may be \emph{less than 4 horizons}. As already discussed in Section~\ref{sec_extreme}, when the physical parameters satisfy the extremality relation ${m^2 + l^2 =  a^2 + e^2 + g^2}$ the two \emph{black-hole horizons ${\HH_b^+, \HH_b^-}$ coincide} because ${r_{+}=r_{-}}$. In such a degenerate case the \emph{extremal horizon is located at}
\begin{equation}
r_b^+ = r_b^-  =  m\,, \label{r-extremerep}
\end{equation}
see (\ref{extremality condition}) and \eqref{r-extreme}, while the two distinct acceleration horizons ${\HH_a^\pm}$ given by (\ref{ra+}) and (\ref{ra-}) remain the same. This is the horizon structure for the family of extremal accelerating Kerr--Newman--NUT black holes, recently studied in \cite{MatejovPodolsky:2021}. If the parameters satisfy ${m^2 + l^2 <  a^2 + e^2 + g^2 }$ the \emph{black-hole horizons ${\HH_b^+, \HH_b^-}$ are absent}. Such hyperextreme spacetimes involve accelerating naked singularities with just two acceleration horizons ${\HH_a^\pm}$.

In the limit ${\alpha \to 0}$ of vanishing acceleration, from (\ref{ra+}), (\ref{ra-}) we formally obtain ${r_a^\pm \to \pm \infty}$ which is consistent with the fact that the two horizons ${\HH_a^\pm}$ \emph{disappear} for non-accelerating Kerr--Newman--NUT black holes. In the complementary limit ${a \to 0}$ of vanishing Kerr-like rotation, we \emph{also} obtain ${r_a^\pm \to \infty}$. This explicitly confirms that there are no accelerating \emph{purely} NUT black holes in the Pleba\'nski--Demia\'nski family of type~D spacetimes. Indeed, by setting ${a=0}$ the metric \eqref{newmetricGP2005} becomes independent of~$\alpha$, and the metric reduces to (\ref{metric-a=0}) representing charged Taub--NUT black holes without acceleration. Nevertheless, accelerating black holes with purely NUT parameter exist \emph{outside} the Pleba\'nski--Demia\'nski family \cite{ChngMannStelea:2006} --- they are of algebraic type~I, and have been recently analyzed in detail in \cite{PodolskyVratny:2020}.

Returning now to the \emph{generic case} with four distinct horizons, it immediately follows from (\ref{r+rep})--(\ref{ra-}) that (assuming non-negative parameters $\alpha$, $a$, and $l$)
\begin{equation}
r_b^-<r_b^+\quad\hbox{always,}\qquad\hbox{while}\quad
r_a^-<r_a^+\quad\hbox{for}\quad  0\le l <a\,. \label{orderhorizons}
\end{equation}
In the limiting case ${l \to a}$ we obtain ${r_a^+ = \alpha^{-1}}$, ${r_a^- \to -\infty}$, while for ${l>a}$ there is ${0<r_a^+<r_a^-}$.

The physically most natural ordering of the horizons
\begin{equation}
r_a^-<r_b^-<r_b^+<r_a^+\,, \label{orderofhorizons}
\end{equation}
in which the two black hole horizons $\HH_b^\pm$ are surrounded by two ``outer'' acceleration horizons $\HH_a^\pm$, requires a \emph{sufficiently small acceleration}. The condition ${r_b^+ < r_a^+}$ explicitly reads
\begin{equation}
\alpha < \frac{1}{r_+}\,\frac{a^2+l^2}{a^2+a\,l} \,, \label{orderofhorizonsexpl}
\end{equation}
while ${r_a^-<r_b^-}$  for any ${0\le l <a}$ because in such a case ${r_a^-<0}$ but ${0<r_b^-}$.

By evaluating $Q$ given by (\ref{newQrep}) at ${r=0}$ we obtain
\begin{equation}
\Q(r=0) =  r_{+}\,r_{-} = a^2-l^2 + e^2 + g^2 >0\qquad\hbox{for}\quad l<a \,. \label{Q(0)}
\end{equation}
Consequently, ${Q>0}$ for any $(r_a^-,r_b^-)$. It follows that the coordinate~$r$ is \emph{spatial} in the regions $(r_a^-,r_b^-)$ and $(r_b^+,r_a^+)$, that is \emph{between} the black-hole and acceleration horizons, while it is temporal in the complementary three regions.

Moreover, using the condition (\ref{orderofhorizonsexpl}) we infer that
\begin{equation}
\frac{\alpha\,a}{a^2+l^2}\, r_{-} (l+a \cos \theta)<
\frac{\alpha\,a}{a^2+l^2}\, r_{+} (l+a \cos \theta)<
\alpha\, r_{+}\,\frac{a^2+a\,l}{a^2+l^2}<1\,. \label{P>0}
\end{equation}
It means that both brackets in the metric coefficient $P(\theta)$ given by (\ref{newP}) are positive, and thus the function $P$ in (\ref{newmetricGP2005}) is \emph{always positive}, retaining the correct signature of the spacetime.

\subsection{Ergoregions}
\label{subsec:ergoregions}

With the rotation parameter $a$, the family of black holes (\ref{newmetricGP2005}) contains \emph{ergoregions} similar to those known from the famous Kerr solution.

The boundary of the ergoregion is defined by the condition ${g_{tt}=0}$, where the corresponding metric coefficient reads
\begin{equation}
g_{tt} = \frac{1}{\Omega^2\rho^2}\, ( P \,a^2\sin^2\theta - Q )\,.
\label{gtt}
\end{equation}
The corresponding condition is thus
\begin{equation}
 Q(r_e) = a^2\sin^2\theta  \,P(\theta) \,,
\label{gtt=0}
\end{equation}
where the metric functions $P(\theta)$ and $Q(r)$ are given by (\ref{newP}) and (\ref{newQ}), respectively. For a fixed value of the angular coordinate $\theta$, the right hand side of (\ref{gtt=0}) is some constant. And since the function $Q(r)$ is of the fourth order, it follows that there are (at most) \emph{four distinct boundaries $r_e$ of the ergoregions} in the direction of $\theta$. These are associated with the corresponding four horizons ${\HH_b^\pm}$ and ${\HH_a^\pm}$, defining the surfaces of infinite redshift, and also the stationary limit at which observers on fixed $r$ and $\theta$ cannot ``stand still''.

\begin{figure}[t]
\centerline{\includegraphics[scale=0.35]{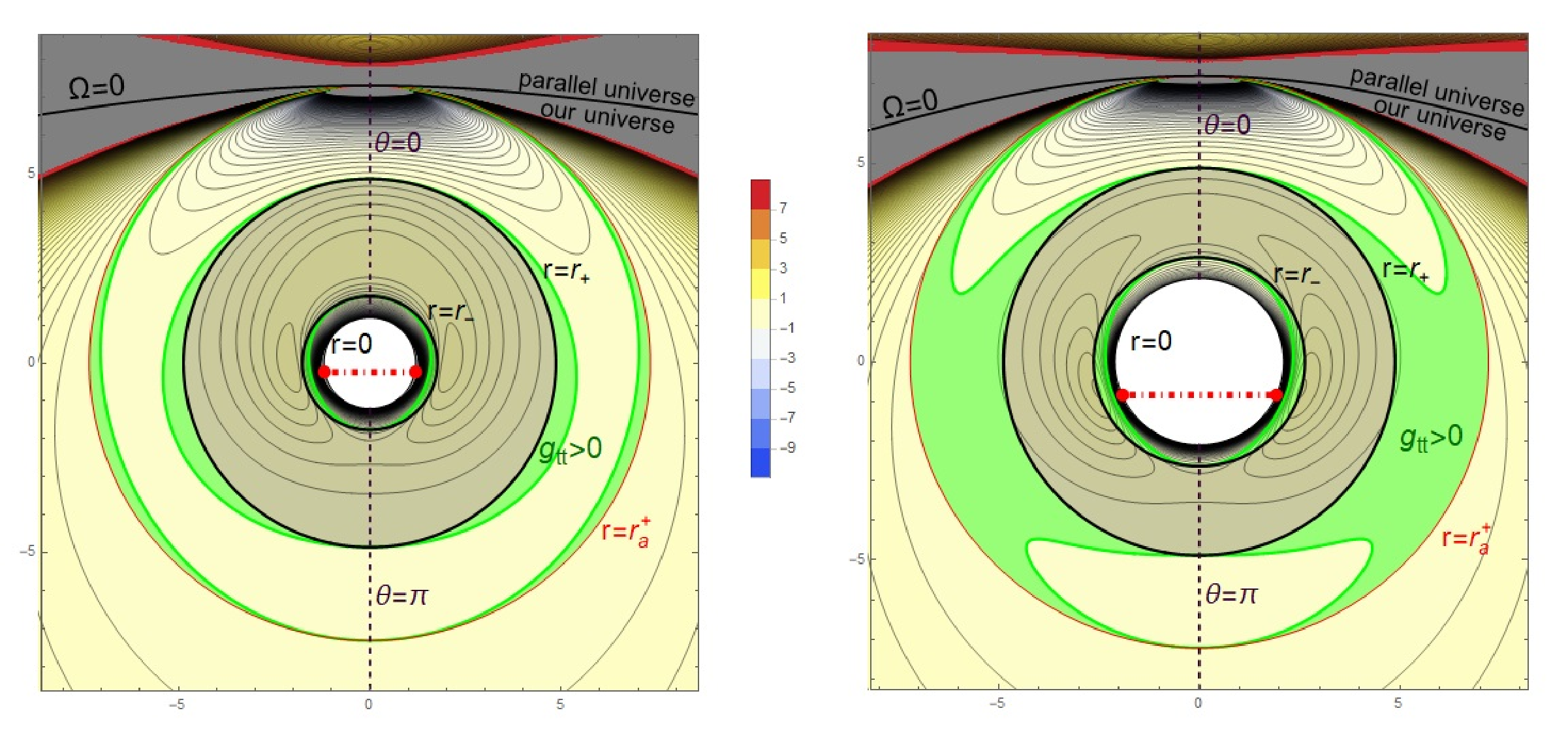}}
\vspace{2mm}
\caption{\small
Plot of the metric function $g_{tt}$ \eqref{gtt} for the accelerating black hole (\ref{newmetricGP2005}) with axes ${\theta = 0}$ and ${\theta = \pi}$. The values of $g_{tt}$ are visualized in quasi-polar coordinates ${{\rm x} \equiv \sqrt{r^2 + (a+l)^2}\,\sin \theta}$, ${\mathrm{y} \equiv \sqrt{r^2 + (a+l)^2}\,\cos \theta\,}$ for ${r \geq 0}$. The grey annulus in the center of each figure localizes the black hole bordered by its horizons $\HH_b^\pm$ at ${r_+}$ and ${r_-}$ (${0<r_-<r_+}$). The acceleration horizon $\HH_a^+$ at ${r_a^+}$ (big red circle) and the conformal infinity $\scri$ at $\Omega=0$ are also shown. The grey curves are contour lines ${g_{tt}(r, \theta)=\hbox{const.}}$, and the values are color-coded from red (positive values) to blue (negative values). The green curves are the isolines ${g_{tt}=0}$ determining the boundary of the ergoregions \eqref{gtt=0} in which ${g_{tt}>0}$ (green regions). They occur close to the horizons near the equatorial plane ${\theta=\pi/2}$. The plot is made for the choice ${m=3}$, ${a=1}$, ${l=0.2}$, ${e=g=1.6}$, ${\alpha=0.12}$ (left) and ${m=3}$, ${a=1.5}$, ${l=0.6}$, ${e=g=1.6}$, ${\alpha=0.12}$ (right). For larger values of $a$ and $l$ the ergoregions are bigger and shifted towards ${\theta = \pi}$. In fact, it can be seen that the ergoregion above the black hole horizon at $r_+$ \emph{is merged} with the ergoregion below the acceleration horizon at $r_a^+$ in the equatorial part.
}
\label{Fig1}
\end{figure}

Solving the equation (\ref{gtt=0}) explicitly is generally complicated but can be plotted using computer, see Fig.~\ref{Fig1}. It is also obvious that \emph{the ergoregion boundary ``touchess'' the corresponding horizon at the poles} ${\theta=0}$ and ${\theta=\pi}$ because there the condition (\ref{gtt=0}) reduces to ${Q(r_e) = 0}$.

In the case of \emph{vanishing acceleration} ${\alpha=0}$, the metric functions (\ref{newP}) and (\ref{newQ}) simplify to ${P=1}$ and ${Q=(r-r_{+})( r-r_{-})}$. Equation (\ref{gtt=0}) reduces to ${ r_e^2 -2m\,r_e + (a^2\cos^2\theta -l^2 + e^2 + g^2)=0}$ which has two roots
\begin{equation}
r_{e\pm}(\theta) = m \pm \sqrt{m^2 + l^2 - e^2 - g^2 - a^2\cos^2\theta }\,.
\label{ergo-alpfa=0}
\end{equation}
This explicitly localizes the two ergoregions for the Kerr--Newman--NUT black holes. As for the standard Kerr black hole, it extends most from the corresponding horizon in the equatorial plane ${\theta=\pi/2}$, in which case $r_{e\pm} = m \pm \sqrt{m^2 + l^2 - e^2 - g^2}$.

On the other hand, \emph{for ${a=0}$ there are no ergoregions} because the condition (\ref{gtt=0}) reduces to ${Q(r_e)=0}$, i.e., the boundaries coincide with the black hole horizons ${\HH_b^\pm}$ at ${r_\pm}$ of the Taub--NUT spacetime (possibly charged). In fact, such horizons become the \emph{Killing horizons} associated with the Killing vector field $\partial_t$, located at ${ m \pm \sqrt{m^2 + l^2 - e^2 - g^2}}$. To summarize, the ergoregions are related only to the Kerr-like rotation represented by the parameter $a$, not to the NUT parameter~$l$. There are no ergoregions in the purely NUT spacetimes.

\subsection{Curvature singularities}
\label{subsec:singularities}

By inspecting the Weyl NP scalar $\Psi_2$ given explicitly by the expression
(\ref{Psi2}) we conclude that the curvature singularities occur if and only if ${\,r+\mathrm{i}\,(l+a \cos \theta)=0}$ (or its complex conjugate). Notice that this complex equation \emph{implies} also ${\rho^2 = r^2+(l+a \cos \theta)^2 = 0}$ which represents the curvature singularity in the Ricci scalar $\Phi_{11}$ given by (\ref{Phi11}) when the electric and magnetic charges $e, g$ are nonzero. Both the real and imaginary parts must vanish, so that the \emph{curvature singularity condition} reads
\begin{equation}
r = 0
\qquad\quad\hbox{and at the same time}\qquad \quad
l+a \cos \theta = 0 \,.  \label{singularity-condition}
\end{equation}

The presence of the curvature singularity is confirmed by the behavior of the Kretschmann scalar ${\mathcal{K} \equiv R_{abcd}\, R^{abcd}}$ given by (\ref{Kretschmann}). The second condition (\ref{singularity-condition}), that is ${L=0}$, implies ${\Omega=1}$, ${\rho^2 = r^2}$, ${F_\pm = r^3}$, ${T_\pm = r^2}$, and
\begin{eqnarray}
K_\pm \rovno m \Big(1 \pm \alpha\,a\,\frac{a^2-l^2}{a^2+l^2} \Big)\,r^3
         \mp l \Big(1 \mp \alpha\,a\,\frac{a^2-l^2+e^2+g^2}{a^2+l^2} \Big)\,r^3
         - (e^2+g^2) \, r^2 \,.  \nonumber
\end{eqnarray}
In the limit ${r\to 0}$ the Kretschmann scalar thus diverges,
\begin{equation}
\mathcal{K} = 48 \, \frac{K_+\,K_-}{r^{12}} \to \infty \,, \label{Kretschmann-r->0}
\end{equation}
because ${K_+\,K_- \sim r^6}$ in the vacuum case, and ${K_+\,K_- \sim r^4}$ in the electrovacuum case.

Now, the important observation is that the necessary (but not sufficient) singularity condition
${l+a \cos \theta = 0}$ \emph{can only be satisfied if} ${|l| \leq |a|}$. Otherwise, the expression ${l+a \cos \theta}$ remains nonzero because ${\cos\theta}$ is bounded to the range $[-1,1]$.

We thus conclude that in the whole family of type~D spacetimes (\ref{newmetricGP2005}) the \emph{curvature singularity structure depends on the relative values of the two twist parameters}, i.e. the Kerr-like rotation $a$ versus the NUT parameter $l$, as follows:
\begin{eqnarray}
l=0\,,\ a=0\,:   &\hbox{singularity at}\ r=0 & \hbox{for  any}\ \theta \,, \nonumber\\
l=0\,,\ a\ne0\,: &\hbox{singularity at}\ r=0 & \hbox{for}\ \theta=\pi/2 \,, \nonumber\\
0<|l|<|a|\,:     &\hbox{singularity at}\ r=0 & \hbox{for}\ \cos \theta = -l/a\,, \label{a-versus-a}\\
l=+a\,:          &\hbox{singularity at}\ r=0 & \hbox{for}\ \theta=\pi\,, \nonumber\\
l=-a\,:          &\hbox{singularity at}\ r=0 & \hbox{for}\ \theta=0\,, \nonumber\\
|l|>|a|>0:       &\hbox{no singularity\,,}   & \nonumber\\
l\ne0\,,\ a=0\,: &\hbox{no singularity\,.}   & \nonumber
\end{eqnarray}

Recall that throughout this paper we naturally assume that all physical parameters ${m, e, g, \alpha, a, l}$ are non-negative. However, for the sake of completeness, in the above table we have admitted the situation in which $a$ and $l$ can be \emph{any real numbers}. In fact, the reflection symmetry ${\varphi \to -\varphi}$ of the metric (\ref{newmetricGP2005}), or equivalently ${t \to -t}$, can be used to change ${a \to -a}$ or ${l \to -l}$ when ${l=0}$ or ${a=0}$, respectively. However, in the generic case when \emph{both} $a$ and $l$ are nontrivial, their relative sign plays the role.

Of course, these results agree with the standard character of the singularity ${r=0}$ of the Schwarzschild, Reissner--Nordstr\"{o}m and (possibly charged) $C$-metric spacetimes (${l=0}$, ${a=0}$), the ring singularity structure of the Kerr and Kerr--Newman black holes (${l=0}$, ${\alpha=0}$), and the absence of curvature singularities in (possibly charged) Taub--NUT spacetime (${a=0}$, ${\alpha=0}$).

\vspace{0mm}
\begin{figure}[t]
\centerline{\includegraphics[scale=0.25]{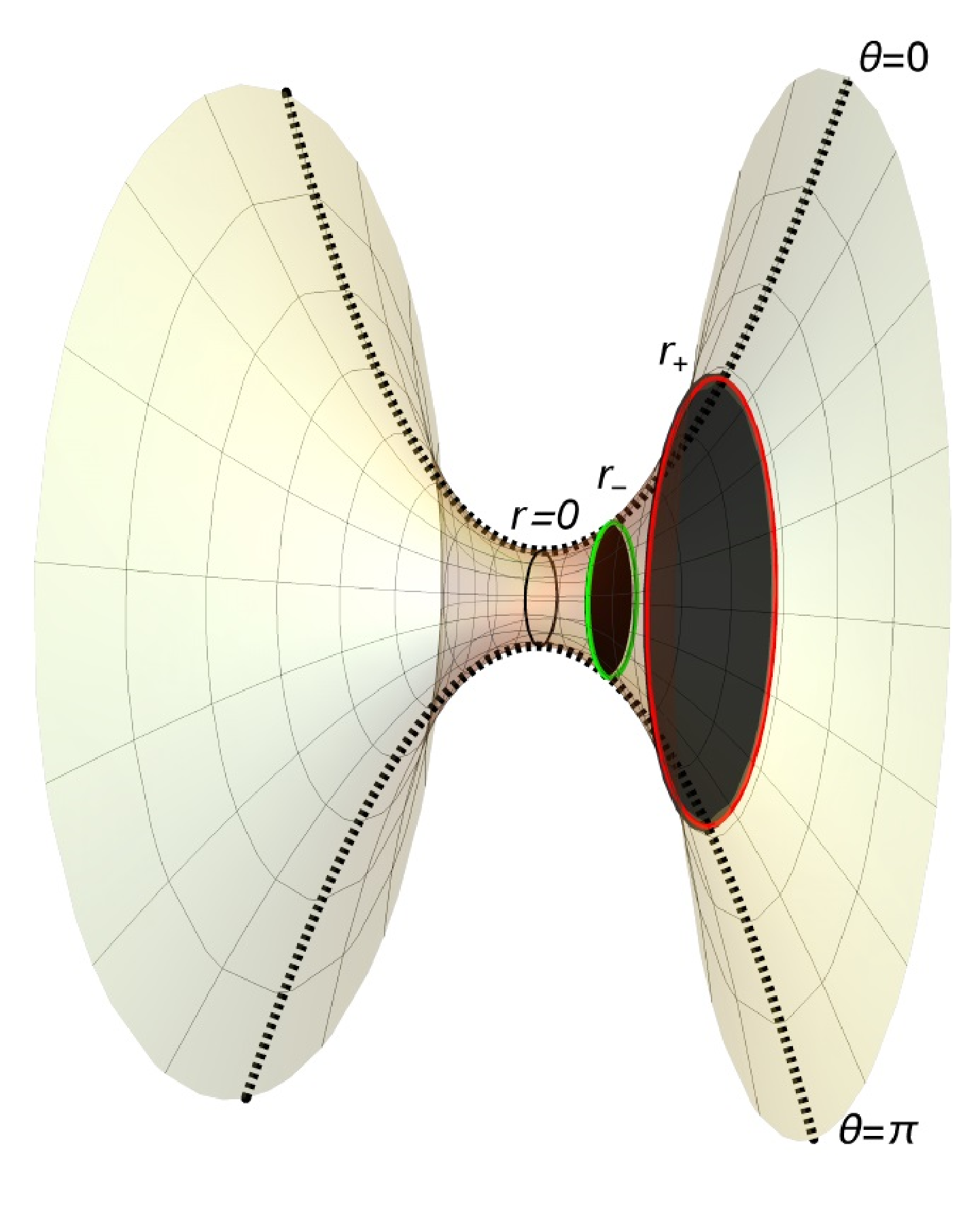}}
\vspace{2mm}
\caption{\small
A schematic visualization of the curvature structure of the generic black hole spacetime (\ref{newmetricGP2005}) using a section with fixed coordinates $t$ and $\varphi$. Away from the singularity located at ${\cos \theta = -l/a,\, r=0}$ it is possible to cross ${r=0}$ from the asymptotically flat universe in the region ${r>0}$ (right part) to another universe in the region ${r<0}$ (left part). In this diagram we also plot the positions of the two black hole horizons ${\HH_b^+}$ and ${\HH_b^-}$ at ${r_{+}}$ and ${r_{-}}$ (red and green circles, respectively), and the
two distinct infinite axes ${\theta = 0}$ and ${\theta = \pi}$ (dashed lines).}
\label{Fig2}
\end{figure}

Finally, it may be useful to graphically represent the global curvature and horizon structure of these black hole spacetimes. On a schematic picture in Fig.~\ref{Fig2} we depict the section ${t=\hbox{const.}}$, ${\varphi=\hbox{const.}}$, taking the full range of ${\theta\in[0,\pi]}$ \emph{distinct} from the specific value ${\cos \theta = -l/a}$. Therefore, the curvature singularity located at ${r=0}$ is \emph{not} encountered, and it is possible to consider the \emph{full range of the coordinate} ${r\in(-\infty,+\infty)}$. In the vicinity of ${r=0}$ the curvature of the spacetime is maximal, in the region ${r>0}$ (the right part of the surface) it decreases to zero, and similarly in the region ${r<0}$ (the left part of the surface) --- far away from the origin the spacetime becomes asymptotically flat. The angular coordinate ${\theta\in[0,\pi]}$ is plotted perpendicularly, completing the full circles ${r=\hbox{const.}}$ (considering also the antipodal section ${\varphi+\pi}$ in the second half of the circle). The resulting ``neck'' or ``wormhole''  connects two distinct universes. Positions of the two black hole horizons ${\HH_b^+}$ at ${r_b^+ \equiv r_{+}}$ and ${\HH_b^-}$ at ${r_b^- \equiv r_{-}}$ are indicated by red and green circles, respectively. Here we assume ${0 < l < a \le \sqrt{a^2 + e^2 + g^2}}$, so that ${0<r_-<r_+}$. In this plot we also show the position of the two distinct full axes ${\theta = 0}$ and ${\theta = \pi}$. These are indicated by dashed lines on top and bottom of the surface.

It should be emphasized that this is only a \emph{schematic picture}, not an embedding and rigorous construction (it cannot be done because the $r$-coordinate is \emph{temporal} between the horizons ${\HH_b^-}$ and ${\HH_b^+}$, and also because the ``point'' ${\cos \theta = -l/a}$, ${r=0}$ is actually the \emph{curvature singularity}.

\vspace{0mm}
\begin{figure}[t]
\centerline{\includegraphics[scale=0.30]{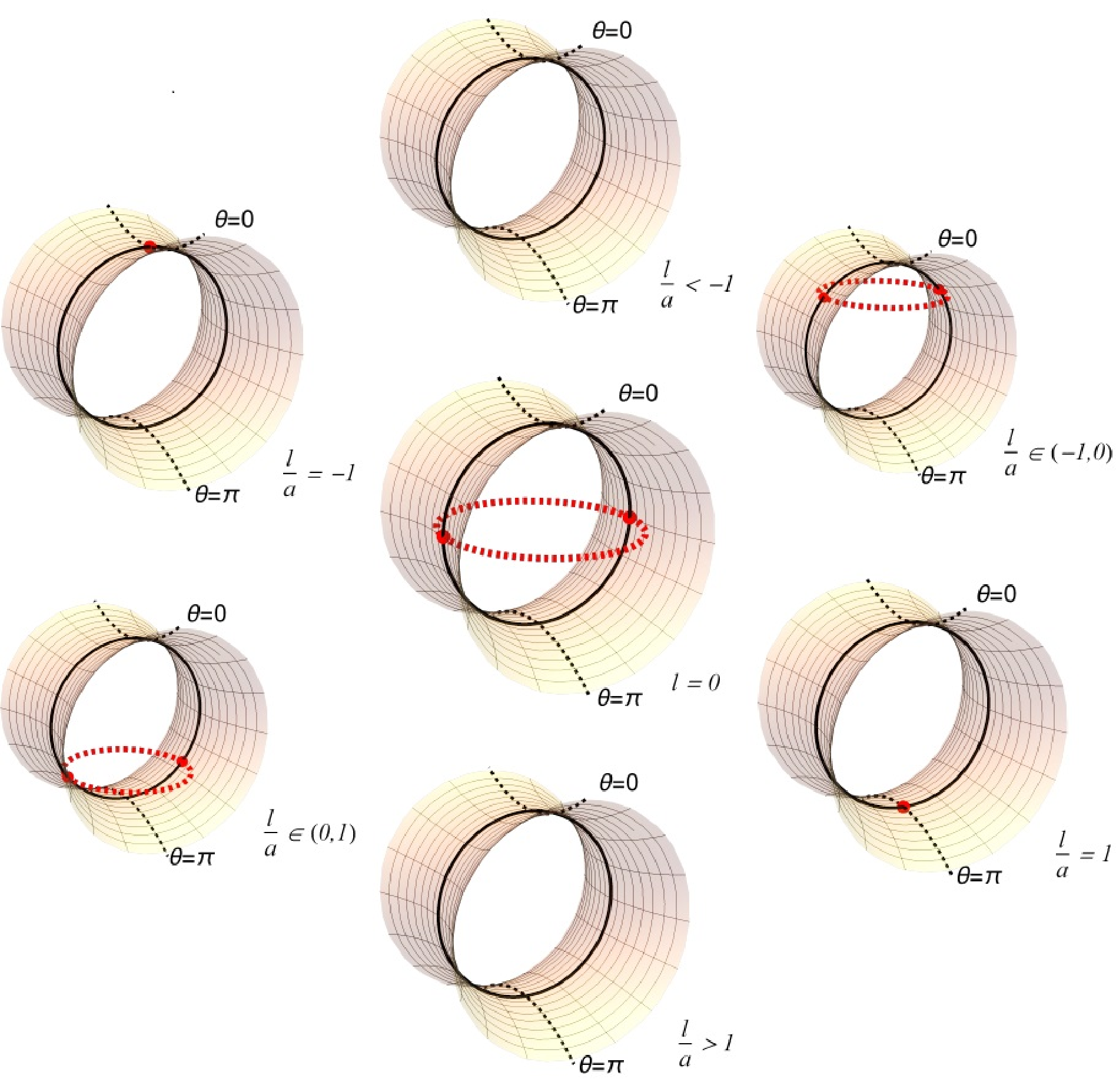}}
\vspace{2mm}
\caption{\small
Schematic visualization of the curvature singularity located at ${r = 0}$, ${\cos \theta = -l/a}$ in the black hole spacetime (\ref{newmetricGP2005}) for 7  distinct choices of the NUT parameter~$l$. For ${|l| \geq |a|}$ such singularity is absent and it is possible to regularly cross ${r=0}$ at \emph{any} $\theta$, entering another asymptotically flat universe.}
\label{Fig3}
\end{figure}

Using the same schematic plot of the central domain of the black hole spacetime, we can also indicate the location of the curvature singularity at ${r=0}$, ${\cos \theta = -l/a}$ for various values of the NUT parameter~$l$ (assuming the same $a$ and other physical parameters). As in Fig.~\ref{Fig2}, the origin  ${r=0}$ is plotted in Fig.~\ref{Fig3} as a black circle around the ``neck'', and the two axes located at ${\theta = 0}$ and ${\theta = \pi}$ are indicated by dashed lines on top and bottom of the surface.

There are 7 such plots in Fig.~\ref{Fig3} corresponding to 7 specific values of~${l/a}$. When the NUT parameter vanishes, ${l=0}$, the curvature singularity is located at ${r=0}$ for ${\theta=\pi/2}$. In the middle plot in Fig.~\ref{Fig2} such a singularity is indicated by \emph{red dots}. In fact, considering also the additional angular coordinate $\varphi\in[0, 2\pi)$, this forms a \emph{ring singularity} of the Kerr--Newman black hole, shown here as the red dashed circle in extra dimension. In the case when ${l=a}$ the curvature singularity is located at the pole ${\theta=\pi}$ (the bottom right plot), while for ${l=-a}$ it is located at the opposite pole ${\theta=0}$ (the top left plot). In the generic case ${|l|<|a|}$, the ring curvature singularity is located at specific $\theta$ between these extremes, such that ${\cos \theta = -l/a}$ (the bottom left and the top right plots). Finally, when ${|l|>|a|}$, there is no curvature singularity (the top and the bottom plots).

In a similar way, by the red dots and the red dashed line we have indicated the position of the ring-like curvature singularity at ${r=0}$ in Fig.~\ref{Fig1}.

\subsection{Conformal diagrams: global structure and infinities}
\label{subsec:global}

In section \ref{subsec:horizon} we have already clarified that the coordinate singularities of the metric located at ${r_b^\pm}$ and ${r_a^\pm}$ correspond to \emph{four} distinct horizons $\HH_b^\pm$ and $\HH_a^\pm$ (provided ${m^2+l^2 \geq a^2+e^2+g^2}$). We will now explicitly construct coordinates which cover the whole spacetime, including these horizons given by the roots ${Q(r)=0}$ of the quartic function \eqref{newQrep}. They will enable us to subsequently derive the corresponding Penrose conformal diagrams showing the global structure of this family of type~D black holes represented by the metric \eqref{newmetricGP2005}.

To this end, we first introduce the \emph{retarded} and \emph{advanced null} coordinates
\begin{eqnarray}
u = t - r_*   \qquad\hbox{and}\qquad   v = t + r_*  \,, \label{uv}
\end{eqnarray}
with the \emph{tortoise} coordinate
\begin{eqnarray}
r_{*} \equiv \int \frac{r^2+(a+l)^2}{Q(r)} \, \dd r \,, \label{tortoise_r}
\end{eqnarray}
and also the corresponding \emph{untwisted angular} coordinates
\begin{eqnarray}
\phi_u \equiv \varphi - a\int\!\frac{\mathrm{d}r}{Q(r)}  \qquad\hbox{and}\qquad   \phi_v \equiv \varphi + a\int\!\frac{\mathrm{d}r}{Q(r)} \,. \label{phiuv}
\end{eqnarray}

Using the \emph{advanced} pair of coordinates $\{v, \phi_v\}$, the metric (\ref{newmetricGP2005}) takes the form
\begin{eqnarray}
\dd s^2 \rovno \frac{1}{\Omega^2}\bigg[\,
\frac{ a^2 P \sin^2 \theta - Q}{\rho^2}\,(\dd v-\T\, \dd \phi_v )^2
+2\,(\dd v-\T\, \dd \phi_v )(\dd r-a\,P \sin^2  \theta \, \dd \phi_v) \nonumber\\
&& \quad +\rho^2 \Big( \frac{\dd \theta^2}{P}+P \sin^2 \theta \, \dd \phi_v^2\Big)\bigg] \,.
\label{metric-advanced}
\end{eqnarray}
The function
\begin{eqnarray}
\T(\theta)    \equiv  a\sin^2\theta +4l\sin^2\!\tfrac{1}{2}\theta \label{T(theta)}
\end{eqnarray}
was introduced to abbreviate the expression. It also enters a useful identity
\begin{eqnarray}
r^2 + (a+l)^2-a\, \T  =  r^2+(l+a \cos \theta)^2 \equiv \rho^2\,. \label{Tidentity}
\end{eqnarray}
Obviously, the metric \eqref{metric-advanced} is regular at ${Q(r)=0}$, so that \emph{the coordinate singularity at the horizons has been removed}.

By employing the complementary \emph{retarded} pair of coordinates $\{u, \phi_u\}$, the metric (\ref{newmetricGP2005}) reads
\begin{eqnarray}
\dd s^2 \rovno \frac{1}{\Omega^2}\bigg[\,
\frac{a^2 P \sin^2 \theta - Q}{\rho^2}\,(\dd u-\T\, \dd \phi_u )^2
-2\,(\dd u-\T\, \dd \phi_u )(\dd r+ a\,P \sin^2  \theta \, \dd \phi_u) \nonumber\\
&& \quad +\rho^2 \Big( \frac{\dd \theta^2}{P}+P \sin^2 \theta \, \dd \phi_u^2\Big)\bigg] \,,
\label{metric-retarded}
\end{eqnarray}
which is also regular at ${Q(r)=0}$.

Actually, these metrics are a considerable generalization of the original coordinate forms of the rotating Kerr--Newman black hole solutions, see Eq.~(1) in \cite{Carter:1968}, Eq.~(5.31) in \cite{HawkingEllis:1973}, or Eq.~(11.4) in \cite{GriffithsPodolsky:2009}. Now it includes not only the usual physical parameters $m, a, e$ (and/or $g$), but also the NUT parameter~$l$ and the acceleration parameter~$\alpha$.

As usual, the next step in construction of the maximal analytic extension of the manifold is to introduce \emph{both the null coordinates~$u$ and~$v$ simultaneously} (dropping $r$ as a coordinate). Clearly, for fixed values of $\phi_v$ and $\theta$ the \emph{radial null geodesics} are simply given by ${v = \text{const.}}$, while for fixed values of $\phi_u$ and $\theta$ the complementary radial null geodesics are given by ${u = \text{const.}}$. Therefore, by employing both the coordinates $u$ and $v$, the causal structure of the spacetime is naturally revealed. Using the relation \eqref{uv} we immediately obtain
\begin{eqnarray}
 v-u = 2\,r_*(r)  \,, \label{r*uv}
\end{eqnarray}
so that
\begin{eqnarray}
 2\,\dd r =  \frac{Q}{r^2+(a+l)^2} \,(\dd v-\dd u) \,. \label{dr*uv}
\end{eqnarray}
This relation can be used to eliminate the $\dd r$-term either from the metric \eqref{metric-advanced} or \eqref{metric-retarded}.

Moreover, due to the simple \emph{factorized} form \eqref{newQrep} of the metric function $Q(r)$, the integral \eqref{tortoise_r} defining the function $r_*(r)$ in \eqref{r*uv} can be calculated explicitly as
\begin{eqnarray}
r_{*}(r) =  k_b^+ \, \log \Big| 1-\frac{r}{r_b^+} \Big| + k_b^- \, \log \Big| 1-\frac{r}{r_b^-} \Big|
       + k_a^+ \, \log \Big| 1-\frac{r}{r_a^+} \Big| + k_a^- \, \log \Big| 1-\frac{r}{r_a^-} \Big| \,,
  \label{r*(r)}
\end{eqnarray}
where the auxiliary constant coefficients are
\begin{eqnarray}
k_b^+ \rovno   \frac{(a^2+l^2)^2\big[r_+^2+(a+l)^2\big]}{2m \big(a^2+l^2+\alpha a (a-l) \, r_+\big) \big(a^2+l^2-\alpha a (a+l) \, r_+\big) }\,,\nonumber \\
k_b^- \rovno - \frac{(a^2+l^2)^2\big[r_-^2+(a+l)^2\big]}{2m \big(a^2+l^2+\alpha a (a-l) \, r_-\big) \big(a^2+l^2-\alpha a (a+l) \, r_-\big) }\,,\nonumber \\
k_a^+ \rovno  -\frac{ (a^2+l^2) \big[(a^2+l^2)^2+\alpha^2 a^2 (a+l)^4  \big]}{2 \, \alpha \,a^2 \big(a^2+l^2-\alpha a (a+l) \, r_+\big) \big(a^2+l^2-\alpha a (a+l) \, r_- \big)}\,, \label{k-hpm}\\
k_a^- \rovno \frac{(a^2+l^2) \big[(a^2+l^2)^2+\alpha^2 a^2 (a^2-l^2)^2 \big]}{2 \, \alpha \,a^2   \big(a^2+l^2+\alpha a (a-l) \, r_+ \big) \big(a^2+l^2+\alpha a (a-l) \, r_-\big)}\,, \nonumber
\end{eqnarray}
each associated with the corresponding horizon $\HH_h^\pm$ located at ${r = r_h^\pm}$, where ${h=b}$ (for the black-hole horizons) or  ${h=a}$ (for the acceleration horizons). Inverting the function \eqref{r*(r)}, we can express the metric functions $Q$, $\rho^2$ and $\Omega^2$ in terms of the null coordinates ${v-u}$ instead of~$r$ by using the relation \eqref{r*uv}.

To obtain the maximal extension of the black-hole manifold represented by \eqref{newmetricGP2005}, we now ``glue together'' different ``coordinate patches'' (charts of the complete atlas) \emph{crossing all the horizons}, until a curvature singularity or conformal infinity (the scri $\scri$) is reached. In order to derive the correct causal structure, it is essential  to employ the null coordinates $u$ and $v$. Therefore, we apply the coordinate patches of the metric form \eqref{metric-advanced} for extending the spacetime across the horizons in the null direction given by the \emph{advanced} coordinate~$v$, while we  apply the coordinate patches of the metric form \eqref{metric-retarded} for extending the spacetime across the horizons in the complementary null direction given by the \emph{retarded} coordinate~$u$. Since both these metrics are regular for~${Q=0}$, the coordinate singularities at \emph{all the horizons}  $\HH_h^\pm$ are removed, step-by-step.

However, to perform this procedure exactly and correctly, two complicated issues must also be clarified. The first problem is the fact, that the distinct coordinate patches \eqref{metric-advanced} and \eqref{metric-retarded} employ  \emph{distinct angular coordinates $\phi_v$ and $\phi_u$}, respectively. The second problem is to prove that thus obtained maximal extension of the manifold is \emph{analytic}.

\newpage
To resolve the first problem associated with distinct angular coordinates $\phi_v$ and $\phi_u$, we can employ the general strategy suggested by Boyer and Lindquist \cite{BoyerLindquist:1967} for the Kerr spacetime and subsequently used also for the charged Kerr--Newman spacetime by Carter \cite{Carter:1968}. The trick is based on using the specific \emph{Killing vector fields} which are \emph{the null generators of the horizons}. In terms of the two coordinate patches \eqref{metric-advanced} and \eqref{metric-retarded}, such special vector fields read
\begin{eqnarray}
\xi^a \equiv \partial_u + \Omega_h \, \partial_{\phi_u}\,,
\qquad\hbox{and also}\qquad
\xi^a \equiv \partial_v + \Omega_h \, \partial_{\phi_v}\,,
\label{Killing}
\end{eqnarray}
where the angular velocity of the given horizon $\HH$ is
\begin{eqnarray}
\Omega_h \rovno \frac{a}{r_h ^2 +(a+l)^2} \,. \label{Omega-h-global}
\end{eqnarray}
Indeed, using the corresponding metric coefficients of \eqref{metric-advanced} and \eqref{metric-retarded}, evaluated at ${Q=0}$, it is straightforward to show that ${\xi^a\xi_a(\HH)=0}$ whenever ${\Omega_h = a/(\,\rho_h^2+ a\,\T\,)}$. Applying the identity \eqref{Tidentity}, we obtain the expression \eqref{Omega-h-global} for \emph{both} the Killing vector fields \eqref{Killing}.

Now, following \cite{Carter:1968, BoyerLindquist:1967} we introduce a special angular coordinate $\phi_h$ \emph{which is constant along the trajectories of both the Killing vector fields}~\eqref{Killing}. Being the  generators of the specific \emph{bifurcate Killing horizon} (a 2-dimensional spatial intersection of the ``advanced'' and the ``retarded'' null horizons), via such new angular coordinate $\phi_h$ a suitable  transition between the corresponding patches is achieved. Technically, it is introduced by the 1-form condition
\begin{eqnarray}
2\,\dd \phi_h \equi \dd \phi_u + \dd \phi_v - \Omega_h (\dd u + \dd v) \,, \label{def-phi_h}
\end{eqnarray}
because ${\dd \phi_h (\xi^a)=0}$ for \emph{both} the Killing vector fields \eqref{Killing}.
Using \eqref{uv} and \eqref{phiuv}, this condition can be integrated to
\begin{eqnarray}
\phi_h = \varphi - \Omega_h \,t \,. \label{def-phi_h-expl}
\end{eqnarray}
Unfortunately, the specific choice of the angular coordinate $\phi_h$ \emph{depends on the given horizon} via its value $r_h$ and thus $\Omega_h$. For this reason,  it is not possible to find a single and simple global coordinate $\phi$ which would conveniently ``cover'' all the four horizons. This drawback was met many years ago already in the Kerr spacetime, so it is not surprising that it reappears in the current context of the complete family of type~D black holes.

An explicit \emph{general} metric form constructed in this way reads
\begin{eqnarray}
\dd s^2 \rovno \frac{1}{4\Omega^2}\bigg[\,
-\frac{Q}{\rho^2}\,\Big((1-\T\Omega_h)(\dd u+\dd v) - 2\T \dd\phi_h\Big)^2
+ Q\rho^2\frac{(\dd u-\dd v)^2 }{[r^2+(a+l)^2]^2}
+ 4\frac{\rho^2}{P}\, \dd \theta^2  \nonumber\\
&& \qquad
+\frac{ P \sin^2 \theta}{\rho^2}\,\Big(\big(a-[r^2+(a+l)^2]\,\Omega_h \big)(\dd u+\dd v) - 2\,[r^2+(a+l)^2]\, \dd\phi_h\Big)^2   \bigg]\,.
\label{metric-doublenull}
\end{eqnarray}
For \emph{non-twisting} black holes without the Kerr-like rotation (${a=0}$) and the NUT parameter (${l=0}$), the metric functions simplify to ${\Omega=1}$, ${P=1}$, ${\rho^2=r^2}$, ${\T=0}$, ${\Omega_h=0}$, so that
\begin{eqnarray}
\dd s^2 \rovno - \frac{Q}{r^2}\,\dd u\,\dd v + r^2(\dd \theta^2 + \sin^2 \theta\,\dd\phi_h^2)\,,
\label{metric-doublenull-nontwist}
\end{eqnarray}
which is the usual form of the spherically symmetric black holes in the double-null coordinates \cite{GriffithsPodolsky:2009}.

On any 2-dimensional section ${\theta=\hbox{const.}}$ and ${\phi_h =\hbox{const.}}$, using \eqref{Omega-h-global}, the general metric \eqref{metric-doublenull}  reduces to
\begin{eqnarray}
\dd \sigma^2 \rovno \frac{1}{4\Omega^2}\bigg[\,
-\frac{(1-\T\Omega_h)^2}{\rho^2}\, Q\,(\dd u+\dd v)^2
+\frac{\rho^2}{[r^2+(a+l)^2]^2}\,Q\,(\dd u-\dd v)^2  \nonumber\\
&& \hspace{8mm}
+\,a^2\,\frac{ P \sin^2 \theta}{\rho^2}\,\frac{(r+r_h)^2(r-r_h)^2}{[r_h^2+(a+l)^2]^2}\,(\dd u+\dd v)^2   \bigg]\,,
\label{metric-doublenull-section}
\end{eqnarray}
which is indeed null at any horizon $r_h$ because ${Q(r_h)=0}$.

Let us now move to the second problem, which is the global extension and investigation of the degree of smoothness (analyticity) of the horizons $\HH_h^\pm$. Restricting ourselves to the sections given by constant values of the angular coordinates~$\theta$ and~$\phi_h$, we introduce \emph{the couples of new null coordinates} $U_{h}^\pm$ and $V_{h}^\pm$, defined as
\begin{eqnarray}
U_{h}^\pm \rovno (-1)^i\, \sign (k_{h}^\pm) \, \exp \Big(\!\! - \!\frac{u}{2k_{h}^\pm}\Big) \,, \label{U}\\
V_{h}^\pm \rovno (-1)^j\, \sign (k_{h}^\pm) \, \exp \Big(\!\! + \!\frac{v}{2k_{h}^\pm}\Big) \,. \label{V}
\end{eqnarray}
Each couple covers the corresponding horizon $\HH_h^\pm$. Moreover, it is characterized by a \emph{particular choice of two integers} ${(i,j)}$ which specify a certain region in the manifold. Generally, there are 5 types of regions which are separated by the four types of horizons $\HH_{h}^\pm$, namely

\begin{eqnarray}
\textbf{Region} & \textbf{Description} & \textbf{Specification of $(i,j)$} \nonumber \\
\hbox{I:}   &\hbox{ asymptotic time-dependent domain between }  \HH_{a}^+ \hbox{ and } \scri & (n-2m+1,n+2m-1)\nonumber\\
\hbox{II:}  &\hbox{ stationary region between }  \HH_{b}^+ \hbox{ and } \HH_{a}^+ & (2n-m,2n+m-1) \nonumber\\
\hbox{III:} &\hbox{ time-dependent domain between the black-hole horizons} & (n-2m,n+2m)   \nonumber\\
\hbox{IV:}  &\hbox{ stationary region between }  \HH_{a}^- \hbox{ and } \HH_{b}^-  & (2n-m+1,2n+m) \nonumber\\
\hbox{V:}   &\hbox{ asymptotic time-dependent domain between }  \scri \hbox{ and } \HH_{a}^-   & (n-2m+1,n+2m-1) \nonumber
\end{eqnarray}
\vspace{0mm}

\noindent
where $m, n$ are arbitrary integers. The corresponding Kruskal--Szekeres-type dimensionless coordinates for every distinct region are
\begin{eqnarray}
T_{h}^\pm = \tfrac{1}{2}(V_{h}^\pm+U_{h}^\pm)\,, \qquad R_{h}^\pm = \tfrac{1}{2}(V_{h}^\pm-U_{h}^\pm) \,.
\end{eqnarray}
Of course, the presence of the \emph{curvature singularity} at ${r=0}$ (implying ${r_*=0}$) for certain values of~$\theta$ restricts the range of the corresponding coordinates $U_{b}^-$ and $V_{b}^-$ in the region~IV to the domain outside ${U_{b}^-V_{b}^- = \pm 1}$.

In terms of these coordinates, the extension across the horizon is regular (in fact, analytic). Indeed, by multiplying and dividing the null coordinates \eqref{U} and \eqref{V} we obtain
\begin{eqnarray}
U_{h}^\pm \, V_{h}^\pm \rovno
     \Big( 1-\frac{r}{r_b^+} \Big)^\frac{k_b^+}{k_{h}^\pm}
     \Big( 1-\frac{r}{r_b^-} \Big)^\frac{k_b^-}{k_{h}^\pm}
     \Big( 1-\frac{r}{r_a^+} \Big)^\frac{k_a^+}{k_{h}^\pm}
     \Big( 1-\frac{r}{r_a^-} \Big)^\frac{k_a^-}{k_{h}^\pm}  \,,  \label{UV}\\
\frac{U_{h}^\pm}{V_{h}^\pm} \rovno (-1)^{i+j}\,
     \exp \Big(\!\! - \!\frac{t}{k_{h}^\pm}\Big)  \,.  \label{U/V}
\end{eqnarray}
The terms ${(\dd u \pm \dd v)^2}$ in the metric \eqref{metric-doublenull-section} become
\begin{eqnarray}
(\dd u \pm \dd v)^2 \rovno \frac{4\,(k_{h}^\pm)^2}{U_{h}^\pm \, V_{h}^\pm}\,
  \bigg(\,\frac{V_{h}^\pm}{U_{h}^\pm}\,(\dd U_{h}^\pm)^2
  \mp 2\, \dd U_{h}^\pm\,\dd V_{h}^\pm
  +\frac{U_{h}^\pm}{V_{h}^\pm}\,(\dd V_{h}^\pm)^2\,\bigg)\,.
\label{du+dv}
\end{eqnarray}
A non-analytic behavior across the horizon $r_h$ may thus occur only at zeros of the product ${U_{h}^\pm \, V_{h}^\pm}$. However, they exactly cancel the zeros of the functions $Q(r)$ in the metric \eqref{metric-doublenull-section}. For example, by choosing the black hole horizon ${r_h=r_b^+\equiv r_+}$, be get ${U_{b}^+ \, V_{b}^+ \propto ( r - r_+)}$ which clearly compensates the corresponding root ${Q \propto ( r - r_+)}$ in \eqref{newQ}. Notice also that the last term in \eqref{metric-doublenull-section} actually vanishes. Therefore, the metric \eqref{metric-doublenull-section} remains finite at ${r_+}$. Of course, the same argument applies to the remaining three horizons.

\newpage
Now we can construct the \emph{Penrose conformal diagrams} which visualize the global structure of the extended manifold. This is achieved by a suitable conformal rescaling of $U_{h}^\pm$ and $V_{h}^\pm$ to the corresponding compactified  null coordinates $\tilde{u}_{h}^\pm$ and $\tilde{v}_{h}^\pm$ defined as
\begin{eqnarray}
\tan \frac{\tilde{u}_{h}^\pm}{2} \equi -\sign (k_{h}^\pm) \, (U_{h}^\pm)^{-\sign (k_{h}^\pm)} = (-1)^{i+1} \exp \Big( \!\! +\! \frac{u}{2 |k_{h}^\pm |}\Big)\,, \\
\tan \frac{\tilde{v}_{h}^\pm}{2} \equi -\sign (k_{h}^\pm) \, (V_{h}^\pm)^{-\sign (k_{h}^\pm)} = (-1)^{j+1} \exp \Big( \!\! -\! \frac{v}{2 |k_{h}^\pm |}\Big) \,.
\end{eqnarray}
Applying the identity ${\,\arctan x + \arctan y  = \arctan(\frac{x+y}{1-xy})\,}$ ${(\hbox{mod}\,\pi)}$ we get
\begin{eqnarray}
\tilde{T}_{h}^\pm \equi \frac{1}{2} (\tilde{v}_{h}^\pm+\tilde{u}_{h}^\pm) =  - \arctan \Bigg[\frac{(-1)^j \exp \Big(\!\!-\!\frac{t+r_*}{2 |k_{h}^\pm |}\Big) + (-1)^i \exp \Big(\frac{t-r_*}{2 |k_{h}^\pm |}\Big)}{1-(-1)^{i+j} \exp \Big(\!\!-\!\frac{r_*}{ |k_{h}^\pm |}\Big)} \Bigg]\,,\\
\tilde{R}_{h}^\pm \equi \frac{1}{2} (\tilde{v}_{h}^\pm-\tilde{u}_{h}^\pm)
 = - \arctan  \Bigg[\frac{(-1)^j \exp \Big(\!\!-\!\frac{t+r_*}{2 |k_{h}^\pm |}\Big) - (-1)^i \exp \Big(\frac{t-r_*}{2 |k_{h}^\pm |}\Big)}{1+(-1)^{i+j} \exp \Big(\!\!-\!\frac{r_*}{ |k_{h}^\pm |}\Big)} \Bigg]\,.
\end{eqnarray}
From these general relations it follows that
\begin{equation}
\tilde{T}_{h}^\pm = \left\{
\begin{array}{ll}
      \vspace{0.5em}(-1)^{j+1} \arctan
      \sdfrac{\cosh \frac{t}{2 |k_{h}^\pm |}}{\sinh \frac{r_*}{2 |k_{h}^\pm |}} &
      \text{ for } i+j \text{ even}\,, \\
      \vspace{0.5em}
      (-1)^{j} \arctan \sdfrac{\sinh \frac{t}{2 |k_{h}^\pm |}}{\cosh \frac{r_*}{2 |k_{h}^\pm |}} &
      \text{ for } i+j \text{ odd, } r_* < 0\,, \\
      (-1)^{j} \arctan \sdfrac{\sinh \frac{t}{2 |k_{h}^\pm |}}{\cosh \frac{r_*}{2 |k_{h}^\pm |}} + \pi &
      \text{ for } i+j \text{ odd, } r_* \geq 0\,, \\
\end{array}
\right.\label{TT}
\end{equation}
and

\begin{equation}
\tilde{R}_{h}^\pm =
\left\{
\begin{array}{ll}
      \vspace{0.5em}(-1)^{j} \arctan \sdfrac{\sinh \frac{t}{2 |k_{h}^\pm |}}{\cosh \frac{r_*}{2 |k_{h}^\pm |}} & \text{ for } i+j \text{ even}\,, \\
      \vspace{0.5em}
      (-1)^{j+1} \arctan \sdfrac{\cosh \frac{t}{2 |k_{h}^\pm |}}{\sinh \frac{r_*}{2 |k_{h}^\pm |}} &
      \text{ for } i+j \text{ odd, } r_*<0\,, \\
     (-1)^{j+1} \arctan \sdfrac{\cosh \frac{t}{2 |k_{h}^\pm |}}{\sinh \frac{r_*}{2 |k_{h}^\pm |}} + \pi &
     \text{ for } i+j \text{ odd, } r_* \geq 0\,. \\
\end{array}
\right.\label{RR}
\end{equation}
Recall that the function ${r_*(r)}$ is given by \eqref{r*(r)} and the coefficients $k_{h}^\pm$ by \eqref{k-hpm}. In particular, the lines of constant~$r$ thus coincide with the lines of constant~$r_*$. Moreover, the condition \eqref{orderofhorizonsexpl} for a reasonably small values of the acceleration parameter~$\alpha$ guarantees that ${k_a^+, k_b^- <0}$ while ${k_a^-, k_b^+ >0}$. Therefore, for every single region the coordinate $r_*$ spans the whole range ${(-\infty,+\infty)}$, and similarly the coordinate~$t$.

\begin{figure}[ht!]
\centerline{\includegraphics[scale=0.245]{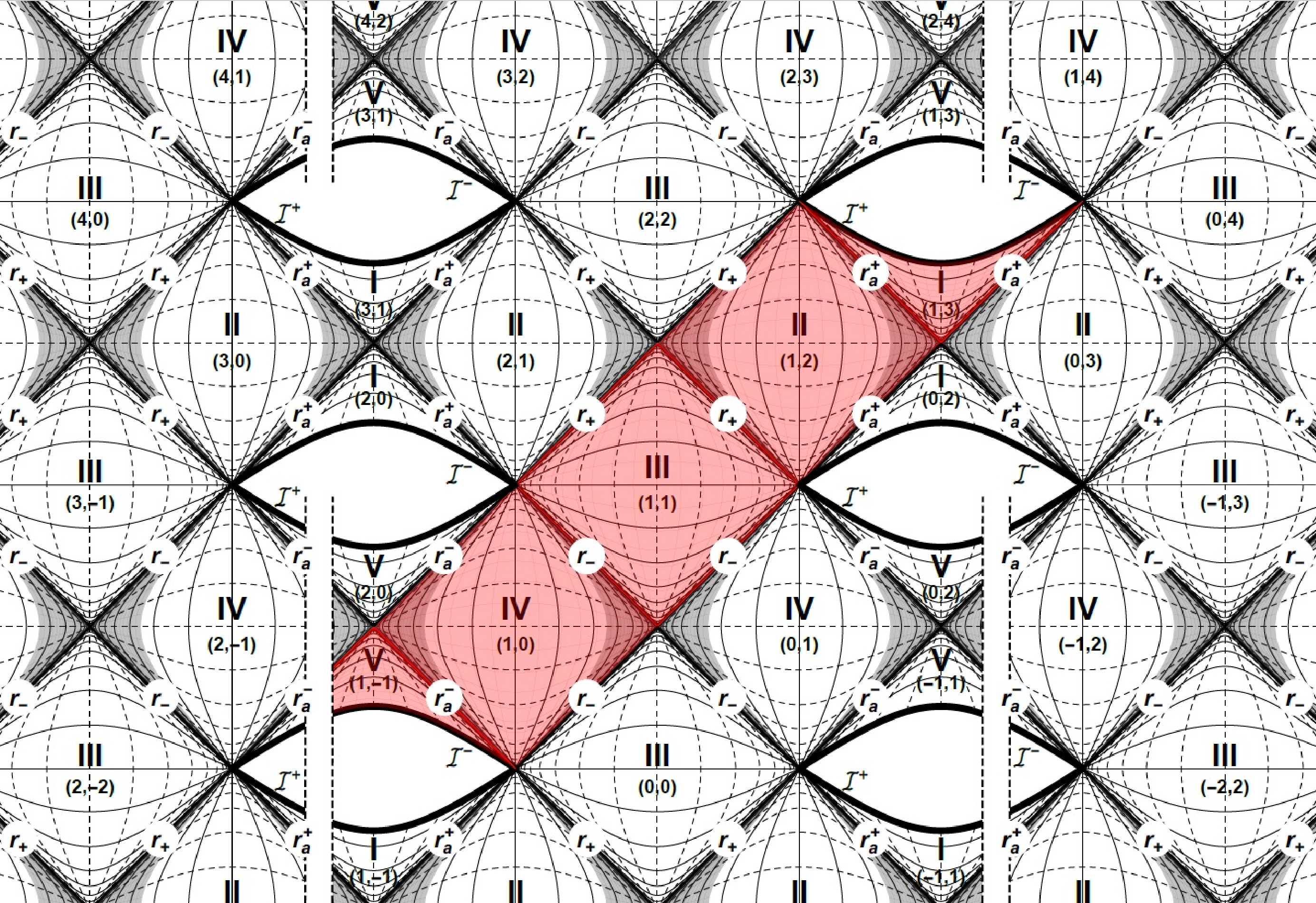}}
\vspace{0mm}
\caption{\small
Penrose conformal diagram of the completely extended spacetime \eqref{newmetricGP2005} showing the global structure of this family of accelerating and rotating charged black holes. We assume the ordering of the four horizons as ${r_a^-<r_-<r_+<r_a^+}$, see \eqref{orderofhorizons}, which occurs for reasonably small  acceleration parameter~$\alpha$, restricted by \eqref{orderofhorizonsexpl}, and  small values of the NUT parameter $l$ such that ${|l|<|a|}$. Here we show a typical 2-dimensional section ${\theta, \phi_h =\hbox{const.}}$ without the curvature singularity at ${r = 0}$, i.e., for any ${\theta=\hbox{const.}}$ such that ${\cos \theta \ne -l/a}$. The double dashed vertical parallel lines indicate a separation of distinct asymptotically flat regions close to $\scri^\pm$ (different ``parallel universes'' that are not necessarily identified). Grey areas in regions~II and~IV close to the horizons denote the ergoregions.
}
\label{Fig4}
\end{figure}

\begin{figure}[ht!]
\centerline{\includegraphics[scale=0.245]{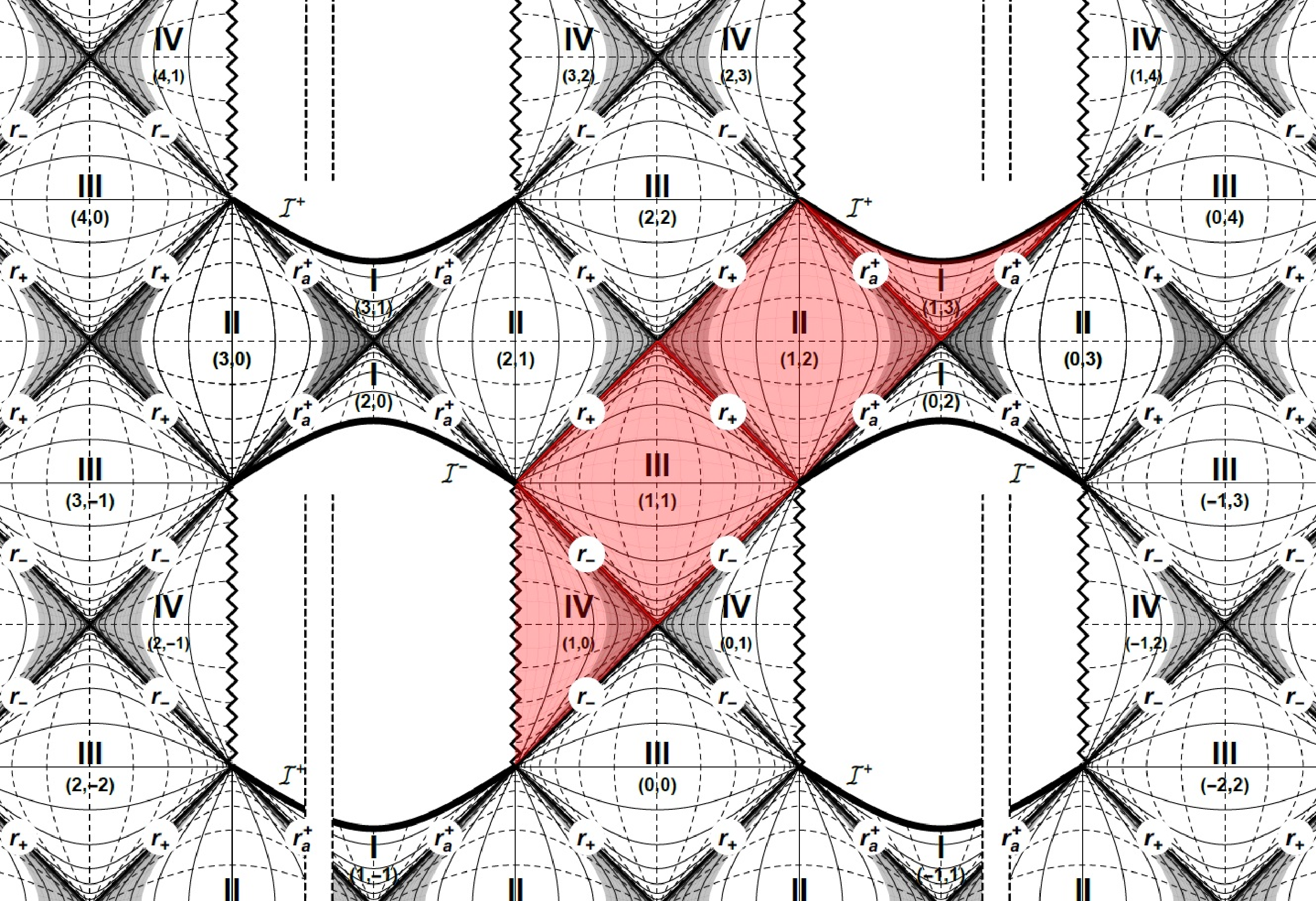}}
\vspace{0mm}
\caption{\small
Penrose conformal diagram of the spacetime \eqref{newmetricGP2005} representing the same black hole as in Fig.~\ref{Fig4} but for the section ${\theta, \phi_h =\hbox{const.}}$ containing the curvature singularity at ${r = 0}$, i.e., for the special value of~$\theta$ such that ${\cos \theta = -l/a}$. In this section, the regions~IV are ``cut in half'' by this singularity at ${r=0}$, so that the acceleration horizon at ${r_a^-<0}$ can not be reached, and the region~V is thus absent.
}
\label{Fig5}
\end{figure}

The explicit relations \eqref{TT}, \eqref{RR} between the compactified coordinates ${\{\tilde{T}_{h}^\pm, \tilde{R}_{h}^\pm \}}$ and the original coordinates ${\{t, r\}}$ of the metric \eqref{newmetricGP2005} for all ${(i,j)}$ can be used for graphical construction of the Penrose diagram which represents the global structure of the extended black-hole manifold, composed of various ``diamond'' regions. The resulting picture is shown in Fig.~\ref{Fig4} and Fig.~\ref{Fig5}.
Fig.~\ref{Fig4} is the Penrose diagram of a \emph{generic} 2-dimensional section through the whole spacetime for any ${\theta=\hbox{const.}}$ such that ${\cos \theta \ne -l/a}$. It \emph{does not contain} the curvature singularity at ${r = 0}$. Fig.~\ref{Fig5} is the complementary Penrose diagram for \emph{the special} value of~$\theta$ such that ${\cos \theta = -l/a}$ which \emph{contains} the curvature singularity at ${r = 0}$ in all its regions~IV (see Sec.~\ref{subsec:singularities} and Fig.~\ref{Fig3}).

\newpage

\vspace{0mm}
\vspace{0mm}

\newpage

It can be seen that the complete manifold consists of an \emph{infinite number of the regions}~I, II, III, IV and V, each identified by the specific pair of integers $(i,j)$. These regions are \emph{separated by the corresponding horizons}. Namely, the regions I and II are separated by the acceleration horizon ${\HH_a^+}$ at $r_a^+$, with the asymptotic region I also bounded by the conformal infinity $\scri$ (the scri) for very large values of $r$. The regions II and III are separated by the black-hole horizon ${\HH_b^+}$ at ${r_b^+\equiv r_+}$, while the regions III and IV are separated by the inner black-hole horizon ${\HH_b^-}$ at ${r_b^-\equiv r_-}$. Finally, the regions IV and V (if present) are separated by the acceleration horizon ${\HH_a^-}$ at $r_a^-$, with the asymptotic region V bounded by the conformal infinity $\scri$ with negative values of $r$. The curves in each region represent the lines of constant~$t$ and~$r$ (dashed or solid, respectively).

In the diagonal \emph{null directions} of these Penrose diagrams we can identify the particular coordinate patches covered by the ``advanced'' metric form \eqref{metric-advanced}, extending from the bottom left $\scri^-$ to the top right $\scri^+$ (for example the pink regions~{I--V} between ${(1,-1)}$ and ${(1,3)}$), and also the complementary ``retarded'' metric form  \eqref{metric-retarded}, extending from the bottom right $\scri^-$ to the top left $\scri^+$ (these are not colored but also contain the regions~{I--V}, for example between ${(-1,1)}$ and ${(3,1)}$). These patches ``share'' the ``central regions''~III (for example ${(1,1)}$). Each of such central region III is bounded  by the inner and outer black-hole horizons at $r_-$ and $r_+$, localizing thus the interior of the corresponding black hole. In the whole extended universe, there are thus \emph{infinitely many black holes} --- they are identified by the regions~III, and labeled by the corresponding specification $(i,j)$, for example $(0,0)$, $(1,1)$, $(2,2)$, $(-2,2)$, $(-1,3)$, $(0,4)$, etc.

Recall that all these black holes are \emph{rotating, NUTed, charged, and accelerating}. Due to their rotation, there are \emph{ergoregions} associated with \emph{all the horizons}, see Sec.~\ref{subsec:ergoregions} and Fig.~\ref{Fig1}. They are represented by the grey areas in the regions~II and~IV close to the horizons.

As shown in Sec.~\ref{subsec:singularities} and schematically depicted in Fig.~\ref{Fig2}, there are two \emph{distinct asymptotically flat universes} associated with each original coordinate patch given by the metric \eqref{newmetricGP2005}, one for ${r\to+\infty}$ and the other for  ${r\to-\infty}$. These can now be identified in the Penrose diagram in Fig.~\ref{Fig4} as the regions~I and~V beyond the acceleration horizons close to $\scri$, respectively. However, the maximal extension has now revealed that \emph{each black hole}, identified by the specific region~III, \emph{is in fact associated with four asymptotically flat regions}, namely the pair of the regions~I and a pair of the regions~V. Two such regions are in the \emph{causal future}, while the remaining two are in the \emph{past}. Moreover, each asymptotically flat region bounded by $\scri$ is \emph{``shared'' by two distinct black holes}.

For example, the \emph{``infinite chain'' of black holes} (regions~III) given by $\ldots$, ${(3,-1)}$, ${(1,1)}$, ${(-1,3)}$, $\ldots$ are located in the ``future universes'' (regions~I) $\ldots$, ${(5,-1)}$, ${(3,1)}$, ${(1,3)}$, ${(-1,5)}$, $\ldots$, while their ``past universes'' (regions~V) are $\ldots$, ${(3,-3)}$, ${(1,-1)}$, ${(-1,1)}$, ${(-3,3)}$, $\ldots$, respectively. However, these ``past universes'' \emph{need not be the same} asymptotically flat regions. Therefore, we inserted the double dashed vertical parallel lines in them to indicate their separation: in general the two regions such as ${(1,-1)}$ are \emph{different} ``causal-past parallel universes'' \emph{with respect to the distinct causal-future universes of the chain of the black holes}. Of course, it is possible to ``artificially'' identify (some of) them --- both the black-hole regions~III and/or their asymptotically flat regions~I and~V. Since there are \emph{infinitely many possibilities of such identifications}, a plethora of various topologically extremely complicated manifolds can be constructed.

Finally, let us remark that the conformal infinities $\scri$ plotted in Figs.~\ref{Fig4},~\ref{Fig5} \emph{does not look null}. This may be surprising because in all the regions~I and~V the spacetime is \emph{asymptotically flat} (excluding the cosmic strings along the axes ${\theta=0}$ and ${\theta=\pi}$, arising as specific topological defects which we will investigate in the next three sections of this paper). Being Minkowski-like, the scri~$\scri$ is indeed null. However, it should be emphasized that the Penrose diagrams in Fig.~\ref{Fig4} and Fig.~\ref{Fig5} are \emph{just 2-dimensional sections} through the global conformal structure of the four-dimensional Lorentzian manifold which is not spherically symmetric. In particular, it turns out that in the presence of acceleration, the \emph{null} conformal infinity $\scri$ of the asymptotically flat regions is indeed represented as the \emph{non-null curve} in the given \emph{section}. This has been thoroughly discussed and analyzed in our previous work on the $C$-metric \cite{GriffithsKrtousPodolsky:2006}, see also Chapter~14 in \cite{GriffithsPodolsky:2009}.

The global extension of the type~D black-hole family of spacetimes obtained in this section seems to be more elegant and also more complete than the preliminary investigation \cite{GriffithsPodolsky:2006a} which employed rather complicated transformations to the Weyl--Lewis--Papapetrou form and subsequently to the boost-rotation-symmetric form of the metric. Moreover, here it is explicitly compactified.

\newpage
\subsection{Cosmic strings (or struts) and deficit angles at ${\theta=0}$ and ${\theta=\pi}$}
\label{subsec:strings}

As shown already in previous works~\cite{GriffithsPodolsky:2005, GriffithsPodolsky:2006}, the metric form \eqref{newmetricGP2005} is convenient for explicit analysis of the \emph{regularity of the poles/axes} located at ${\theta=0}$ and ${\theta=\pi}$, respectively. This is now further improved with the new metric functions \eqref{newOmega}--\eqref{newQ}.

The spatial axes of symmetry are associated with the Killing vector field $\partial_\varphi$, identified as its degenerate points. These are located at the coordinate singularities of the function $\sin\theta$ in the metric (\ref{newmetricGP2005}) which appear at ${\theta=0}$ and ${\theta=\pi}$. Therefore, the range of the spatial coordinate~$\theta$ must be constrained to ${\theta\in[0,\pi]}$.

Recall that there are six physical parameters in the new metric \eqref{newmetricGP2005}, namely $m, a, l, \alpha, e, g$, which represent mass, Kerr-like rotation, NUT parameter, acceleration, electric and magnetic charges of the black hole, respectively. However, there is also the \emph{seventh free parameter --- the conicity $C$ hidden in the range of the angular coordinate}
\begin{equation}
\varphi\in[0,2\pi C)\,,
 \label{conicity}
\end{equation}
which has not yet been specified. We will demonstrate its physical meaning by relating it to the \emph{deficit (or excess) angles} of the \emph{cosmic strings (or struts)}. Their tension is the \emph{physical source of the acceleration of the black holes}. These are basically topological defects associated with \emph{conical singularities} around the two distinct axes. In addition, for nonvanishing NUT parameter~$l$ these cosmic strings or struts are \emph{rotating}, thus introducing specific internal twist to the entire spacetime. We will now analyze them in more detail.

Let us start with investigation of the (non)regularity of the \emph{first axis of symmetry} ${\theta=0}$ in the metric (\ref{newmetricGP2005}). Consider a small circle around it given by ${\theta=\hbox{const.}}$, with the range of $\varphi$ given by \eqref{conicity}, assuming fixed $t$ and~$r$. The invariant length of its \emph{circumference} is ${\int_0^{2\pi C}\!\! \sqrt{g_{\varphi\varphi}}\, \dd\varphi}$, while its \emph{radius} is ${\int_0^{\theta}\! \sqrt{g_{\theta\theta}}\, \dd\theta}$. The axis is regular if their fraction in the limit ${\theta\to 0}$  is equal to ${2\pi}$. However, in general we obtain
\begin{equation}
 f_0 \equiv \lim_{\theta\to0} \frac{\hbox{circumference}}{\hbox{radius}}
 =\lim_{\theta\to0} \frac{2\pi C \sqrt{g_{\varphi\varphi}} }{ \theta\,\sqrt{g_{\theta\theta}}}  \,.
 \label{Accel-con0}
\end{equation}
For the metric \eqref{newmetricGP2005}, the relevant metric functions are
\begin{equation}
 g_{\varphi\varphi} = \frac{1}{\Omega^2\rho^2}\,
   \Big[\, P \big(r^2+(a+l)^2\big)^2\sin^2\!\theta
   - Q \big(a\sin^2\theta +4l\sin^2\!\tfrac{1}{2}\theta \,\big)^2\,\Big]\,,\qquad
 g_{\theta\theta} = \frac{\rho^2}{\Omega^2 P}  \,.
 \label{gphiphi-gthetatheta0}
\end{equation}
For very small values of $\theta$, the second term in $g_{\varphi\varphi}$ proportional to $Q$ becomes negligible with respect to the first term proportional to $P$, so that we obtain ${ g_{\varphi\varphi} \approx P \big(r^2+(a+l)^2\big)^2\,\theta^2/\Omega^2\rho^2}$.
Straightforward evaluation of the limit (\ref{Accel-con0}) gives
\begin{equation}
 f_0 =  2\pi C\,P(0) = 2\pi C\,
 \Big( 1-\alpha\,\frac{a^2+al}{a^2+l^2}\, r_{+} \Big)\Big( 1-\alpha\,\frac{a^2+al}{a^2+l^2}\, r_{-} \Big)\,.
 \label{f0}
\end{equation}
\emph{The axis ${\theta=0}$ in the metric (\ref{newmetricGP2005}) can thus be made regular by the unique choice}
\begin{eqnarray}
 C= C_0 \equi
\Big[ \Big( 1-\alpha\,\frac{a^2+al}{a^2+l^2}\, r_{+} \Big)\Big( 1-\alpha\,\frac{a^2+al}{a^2+l^2}\, r_{-} \Big)\Big]^{-1}
 \label{C0}\\
 \rovno \Big[ 1 - 2\alpha m\,\frac{a^2+al}{a^2+l^2} + \alpha^2\,\Big(\frac{a^2+al}{a^2+l^2}\Big)^2\, (a^2-l^2+e^2+g^2)\Big]^{-1} , \nonumber
\end{eqnarray}
where we have employed the relations \eqref{r+}, \eqref{r-}. Notice that for vanishing acceleration $\alpha$, this regularization condition is simply ${C_0=1}$.

Analogously, it is possible to regularize the \emph{second axis of symmetry} ${\theta=\pi}$. Now, the conceptual problem is that the metric function $g_{\varphi\varphi}$ in (\ref{gphiphi-gthetatheta0}), and thus the circumference, does \emph{not} approach zero in the limit ${\theta\to\pi}$ due to the presence of the term ${4l\sin^2\!\tfrac{1}{2}\theta }$. This problem can be resolved by the same procedure as for the classic Taub--NUT solution (see the transition between the metrics (12.1) and (12.3) in \cite{GriffithsPodolsky:2009}), namely by applying the transformation of the time coordinate
\begin{equation}
t_{\pi} \equiv t - 4l\,\varphi\,.
 \label{t-tpi}
\end{equation}
The metric (\ref{newmetricGP2005}) then becomes
\begin{align}
\dd s^2 = \frac{1}{\Omega^2} &
  \left(-\frac{Q}{\rho^2}\left[\,\dd t_{\pi}- \left(a\sin^2\theta - 4l\cos^2\!\tfrac{1}{2}\theta \right)\dd\varphi \right]^2
   + \frac{\rho^2}{Q}\,\dd r^2 \right. \nonumber\\
& \quad \left. + \,\frac{\rho^2}{P}\,\dd\theta^2
  + \frac{P}{\rho^2}\,\sin^2\theta\, \big[ a\,\dd t_{\pi} -\big(r^2+(a-l)^2 \big)\,\dd\varphi \big]^2
 \right), \label{newmetricGP2005-Regular-pi}
\end{align}
i.e.,
\begin{equation}
 g_{\varphi\varphi} = \frac{1}{\Omega^2\rho^2}\,
   \Big[\, P \big(r^2+(a-l)^2\big)^2\sin^2\!\theta
   - Q \big(a\sin^2\theta -4l\cos^2\!\tfrac{1}{2}\theta \,\big)^2\,\Big]\,,\qquad
 g_{\theta\theta} = \frac{\rho^2}{\Omega^2 P}  \,.
 \label{gphiphi-gthetatheta-pi}
\end{equation}
Thus, for ${\theta \to \pi}$ we get ${ g_{\varphi\varphi} \approx P \big(r^2+(a-l)^2\big)^2\,(\pi-\theta)^2/\Omega^2\rho^2}$.
The radius of a small circle around the axis ${\theta=\pi}$  is ${\int_{\theta}^{\pi}\! \sqrt{g_{\theta\theta}}\,\dd\theta}$, so that the  fraction
\begin{equation}
 f_\pi \equiv \lim_{\theta\to \pi} \frac{\hbox{circumference}}{\hbox{radius}}
 =\lim_{\theta\to\pi} \frac{2\pi C \sqrt{g_{\varphi\varphi}} }{ (\pi-\theta)\,\sqrt{g_{\theta\theta}}}  \,,
 \label{Accel-conpi}
\end{equation}
is
\begin{equation}
 f_\pi =  2\pi C\,P(\pi) = 2\pi C\,
 \Big( 1+\alpha\,\frac{a^2-al}{a^2+l^2}\, r_{+} \Big)\Big( 1+\alpha\,\frac{a^2-al}{a^2+l^2}\, r_{-} \Big)\,.
 \label{fpi}
\end{equation}
\emph{The axis ${\theta=\pi}$ in the metric (\ref{newmetricGP2005-Regular-pi}) can thus be made regular by the unique choice}
\begin{eqnarray}
 C= C_\pi \equi
\Big[ \Big( 1+\alpha\,\frac{a^2-al}{a^2+l^2}\, r_{+} \Big)\Big( 1+\alpha\,\frac{a^2-al}{a^2+l^2}\, r_{-} \Big)\Big]^{-1}
 \label{Cpi}\\
 \rovno \Big[ 1 + 2\alpha m\,\frac{a^2-al}{a^2+l^2} + \alpha^2\,\Big(\frac{a^2-al}{a^2+l^2}\Big)^2\, (a^2-l^2+e^2+g^2)\Big]^{-1} . \nonumber
\end{eqnarray}
With such a choice, there is a \emph{deficit angle} $\delta_0$ (conical singularity) \emph{along the first axis} ${\theta=0}$, namely
\begin{eqnarray}
 \delta_0 \equi 2\pi-f_0  \nonumber \\
  \rovno 8\pi\, \alpha \, \frac{ a^2\,[m(a^2+l^2)-\alpha a l (a^2-l^2+e^2+g^2)] } { (a^2+l^2)^2 + 2\alpha m (a^2-al)(a^2+l^2)+\alpha^2 (a^2-al)^2 (a^2-l^2+e^2+g^2) } \,.
 \label{delta0}
\end{eqnarray}
For black holes without the NUT parameter (${l=0}$) this expression simplifies to
\begin{equation}
 \delta_0 =  \frac{ 8\pi\, \alpha \,m } { 1 + 2\alpha m +\alpha^2 (a^2+e^2+g^2) } \,,
 \label{delta0for-l=0}
\end{equation}
recovering the previous results for rotating charged  $C$-metric, see Chapter~14 in \cite{GriffithsPodolsky:2009}. The tension in the \emph{cosmic string along ${\theta=0}$ pulls the black hole, causing its uniform acceleration}. Such a string \emph{extends to the full range of the radial coordinate} ${r\in (-\infty,+\infty)}$, connecting ``our universe'' with the ``parallel universe'' through the nonsingular black-hole interior close to ${r=0}$.

Complementarily, when the first axis of symmetry ${\theta=0}$ is made regular by the choice (\ref{C0}), there is necessarily an \emph{excess angle} $\delta_\pi$ along the second axis ${\theta=\pi}$, namely
\begin{eqnarray}
 \delta_\pi \equi 2\pi-f_\pi \nonumber \\
  \rovno -8\pi\, \alpha \, \frac{ a^2\,[m(a^2+l^2)-\alpha a l (a^2-l^2+e^2+g^2)] }
   { (a^2+l^2)^2 - 2\alpha m (a^2+al)(a^2+l^2)+\alpha^2 (a^2+al)^2 (a^2-l^2+e^2+g^2) }   \,,
 \label{deltapi}
\end{eqnarray}
which simplifies to
\begin{equation}
 \delta_\pi =  -\frac{ 8\pi\, \alpha \,m } { 1 - 2\alpha m +\alpha^2 (a^2+e^2+g^2) } \,,
 \label{deltapifor-l=0}
\end{equation}
for ${l=0}$. As in the $C$-metric, this represents the \emph{cosmic strut located along ${\theta=\pi}$ between the pair of black holes, pushing them away from each other} in opposite spatial directions.

We observe that ${\delta_0 = 0 = \delta_\pi}$ whenever ${\alpha=0}$. In such a case \emph{both the axes are regular}, there is no physical cause of the acceleration and the Kerr--Newman--NUT black holes do not move.

Interestingly, both the axes ${\theta=0}$ and ${\theta=\pi}$ can be \emph{simultaneously regular even for non-vanishing acceleration}~$\alpha$ when all six physical parameters satisfy the special constraint
\begin{equation}
 m(a^2+l^2) = \alpha a l\, (a^2-l^2+e^2+g^2)  \,.
 \label{bothregularaxes}
\end{equation}
The non-trivial constraint requires \emph{both} ${a\ne0}$ and ${l\ne0}$. Actually, this is a nice compact form of the condition given on page 313 of \cite{GriffithsPodolsky:2009}, when the relations \eqref{PD-par-trans} for the physical parameters and also the convenient gauge choice \eqref{ChoiceOmega-al} are employed. This again demonstrates the advantages of the new form of the metric \eqref{newmetricGP2005}.

However, the condition  \eqref{bothregularaxes} is \emph{not satisfied for small values of the acceleration}~$\alpha$ obeying the inequality \eqref{orderofhorizonsexpl} which guarantees the natural ordering of the four horizons \eqref{orderofhorizons}. Indeed, \eqref{bothregularaxes} can be rewritten as ${m(a^2+l^2) = \alpha a l\, r_{+}\,r_{-}}$. Now applying \eqref{orderofhorizonsexpl}, and assuming $m,a,l$ \emph{all positive}, we get the relation
\begin{equation}
 m < \frac{l}{a+l}\,r_{-} < r_{-}   \,.
 \label{INEQbothregularaxes}
\end{equation}
It is in clear contradiction with \eqref{r-} which implies ${m > r_{-}}$.

\subsection{Rotation of the cosmic strings (or struts)}
\label{subsec:rotatingstrings}

With a generic NUT parameter~$l$, the \emph{cosmic strings (or struts) are rotating}. This can be seen by calculating the \emph{angular velocity} parameter~$\omega_\theta$ of the metric, see~\cite{ChngMannStelea:2006}, along the two different axes ${\theta=0}$ and ${\theta=\pi}$, namely
\begin{equation}
 \omega_\theta \equiv\frac{g_{t\varphi}}{g_{tt}}  \,.
 \label{omega}
\end{equation}

For the general form of the new metric (\ref{newmetricGP2005}), where
\begin{align}
 g_{t\varphi} &= \frac{1}{\Omega^2\rho^2}\,
   \Big[\, Q \big(a\sin^2\theta +4l\sin^2\!\tfrac{1}{2}\theta \,\big)
   - a \big(r^2+(a+l)^2\big)P\sin^2\!\theta\Big]  \,,\nonumber \\
 g_{tt} &= \frac{-1}{\Omega^2 \rho^2}\,
   \Big[\, Q - a^2 P \sin^2 \theta\,\Big]\,,
 \label{gtt-gtphi0}
\end{align}
we obtain
\begin{equation}
 \omega_\theta = -\frac{Q \big(a\sin^2\theta +4l\sin^2\!\tfrac{1}{2}\theta \,\big)
   - a \big(r^2+(a+l)^2\big)P\sin^2\!\theta}
   {Q - a^2 P \sin^2 \theta}\,.
 \label{omega-0}
\end{equation}
Now we take any fixed value of $r$ away from the horizons, so that ${Q\ne0}$ is a non-vanishing constant. Then the limits ${\theta\to0}$ and ${\theta\to\pi}$ are
\begin{equation}
 \omega_0  = 0 \qquad\hbox{and}\qquad  \omega_\pi  = -4l  \,,  \label{omega0pi}
\end{equation}
respectively. The first axis ${\theta=0}$ is thus \emph{non-rotating}, while the second axis ${\theta=\pi}$ rotates and its \emph{angular velocity is directly and solely determined by the NUT parameter~$l$}. Notice that $\omega_\pi$ is independent of the Kerr-like parameter~$a$, and it also does not depend on the conicity parameter~$C$. The rotational character of the axis is thus a specific feature determined by the NUT parameter~$l$, which is clearly \emph{independent} of the possible deficit angles defining the cosmic string/strut along the same axis.

By changing the time coordinate as \eqref{t-tpi}, we obtain the alternative metric \eqref{newmetricGP2005-Regular-pi} for which
\begin{align}
 g_{t_{\pi}\varphi} &= \frac{1}{\Omega^2\rho^2}\,
   \Big[\, Q \big(a\sin^2\theta -4l\cos^2\!\tfrac{1}{2}\theta \,\big)
   - a \big(r^2+(a-l)^2\big)P\sin^2\!\theta\Big]  \,,\nonumber\\
 g_{t_{\pi}t_{\pi}} &= \frac{-1}{\Omega^2 \rho^2}\,
   \Big[\, Q - a^2 P \sin^2 \theta\,\Big]\,,
 \label{gtt-gtphi-pi}
\end{align}
so that
\begin{equation}
 \omega_\theta = -\frac{Q \big(a\sin^2\theta -4l\cos^2\!\tfrac{1}{2}\theta \,\big)
   - a \big(r^2+(a-l)^2\big)P\sin^2\!\theta}
   {Q - a^2 P \sin^2 \theta}\,.
 \label{omega-pi}
\end{equation}
The corresponding angular velocities of the two axes are thus
\begin{equation}
 \omega_0 = 4l \qquad\hbox{and}\qquad  \omega_\pi  = 0  \,.  \label{omega0piregpi}
\end{equation}
In this case, the situation is complementary to \eqref{omega0pi}: the axis ${\theta=0}$ rotates, while the axis ${\theta=\pi}$
is non-rotating.

It is interesting to observe that there is a \emph{constant difference} ${\Delta\omega\equiv\omega_0-\omega_\pi=4l}$ between the angular velocities of the two rotating cosmic strings or struts, directly given by the NUT parameter~$l$ (irrespective of the value of $a$ or the choice of $C$). The NUT parameter is thus responsible for the \emph{difference} between the magnitude of rotation of the two axes ${\theta=0}$ and ${\theta=\pi}$.

\subsection{Closed timelike curves around the rotating strings (or struts)}
\label{subsec:CTCrotatingstrings}

In the vicinity of the rotating cosmic strings or struts located along ${\theta=0}$ or ${\theta=\pi}$, the black-hole spacetime with twist can serve as a specific time machine because (as in the classic Taub--NUT solution) there are \emph{closed timelike curves}.

To identify these pathological causality-violating regions we will consider simple curves in the spacetime, namely \emph{circles around the axes of symmetry} ${\theta=0}$ or ${\theta=\pi}$ such that only the \emph{periodic} angular coordinate ${\varphi\in[0,2\pi C)}$ changes, while the remaining coordinates $t$,~$r$ and~$\theta$ are kept fixed. The corresponding tangent (velocity) vectors are thus proportional to the \emph{Killing vector field}~$\partial_\varphi$. Its norm is determined just by the metric coefficient $g_{\varphi\varphi}$, which for the general metric \eqref{newmetricGP2005} has the form \eqref{gphiphi-gthetatheta0}. There exist regions
such that ${g_{\varphi\varphi}<0}$, where the circles (orbits of the axial symmetry) are \emph{closed timelike curves}. These pathological regions are explicitly given by the condition
\begin{equation}
P(\theta) \big(r^2+(a+l)^2\big)^2\sin^2\!\theta   <   Q(r) \big(a\sin^2\theta +4l\sin^2\!\tfrac{1}{2}\theta \,\big)^2 \,,
 \label{gphiphi-general-condition}
\end{equation}
where the functions $P(\theta)$, $Q(r)$ are given by \eqref{newP}, \eqref{newQ}. In particular, for ${l=0}$, ${g=0}$, ${\alpha=0}$ this reduces to ${\,r^2+a^2+\rho^{-2}\,(2mr-e^2)\,a^2\sin^2\theta<0\,}$ which is exactly the condition (27) derived in \cite{Carter:1968} for the Kerr--Newman family of black holes.

\begin{figure}[t]
\centerline{\includegraphics[scale=0.35]{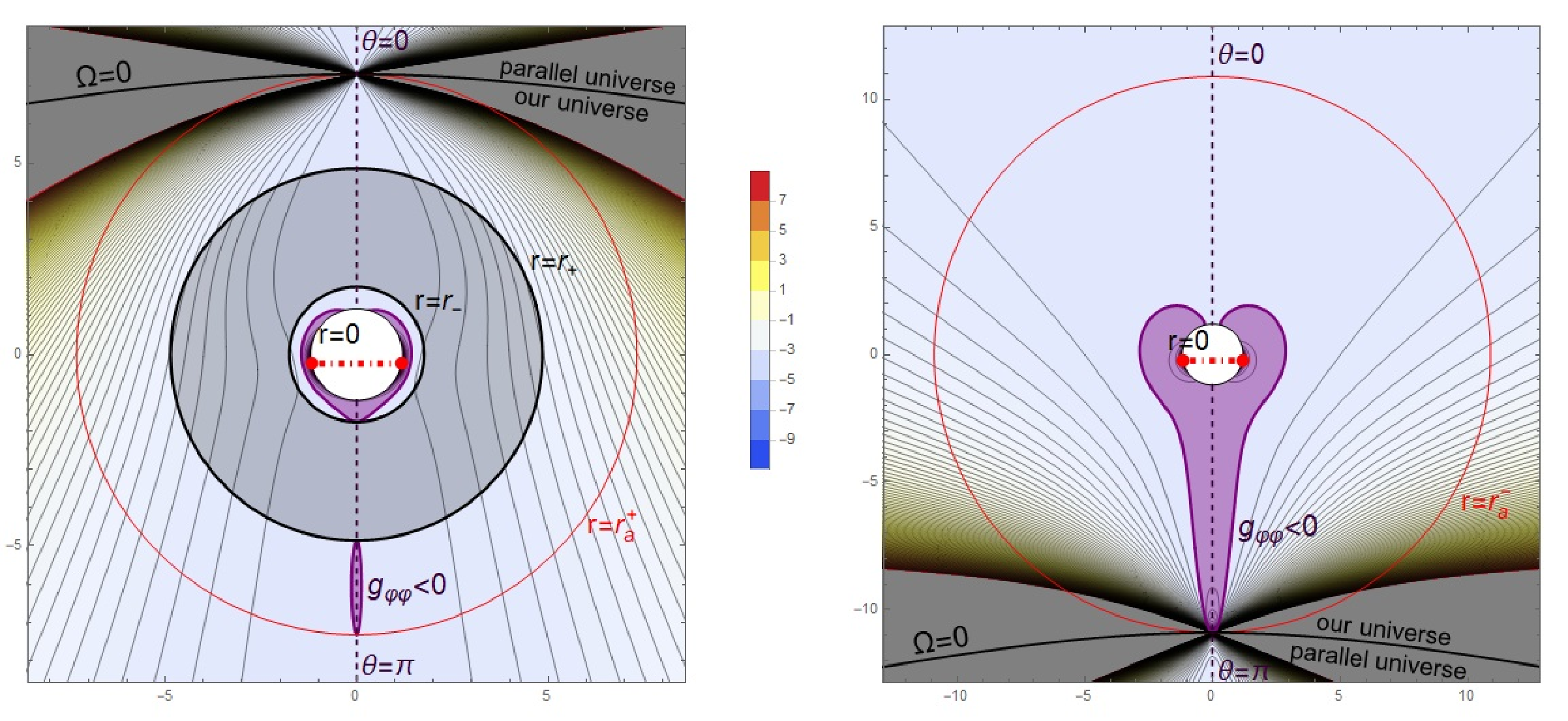}}
\vspace{2mm}
\caption{\small
Plot of the metric function $g_{\varphi\varphi}$ \eqref{gphiphi-gthetatheta0} for the accelerating black hole (\ref{newmetricGP2005}) with a regular axis ${\theta = 0}$ and rotating cosmic string along  ${\theta = \pi}$. The values of $g_{\varphi\varphi}$ are visualized in quasi-polar coordinates ${{\rm x} \equiv \sqrt{r^2 + (a+l)^2}\,\sin \theta}$, ${\mathrm{y} \equiv \sqrt{r^2 + (a+l)^2}\,\cos \theta\,}$ for ${r \geq 0}$ (left) and ${r\leq 0}$ (right). The grey annulus in the center of the left figure localizes the black hole bordered by its horizons $\HH_b^\pm$ at ${r_+}$ and ${r_-}$ (${0<r_-<r_+}$). The acceleration horizons $\HH_a^\pm$ at ${r_a^+}$ and ${r_a^-}$ (big red circles) and the conformal infinity $\scri$ at $\Omega=0$ are also shown. The grey curves are contour lines ${g_{\varphi\varphi}(r, \theta)=\hbox{const.}}$, and the values are color-coded from red (positive values) to blue (negative values); extremely large values are cut. The purple curves are the isolines ${g_{\varphi\varphi}=0}$ determining the boundary of the pathological regions (\ref{gphiphi-general-condition}) with closed timelike curves. They occur close to the axis ${\theta=\pi}$ (purple regions where ${g_{\varphi\varphi}<0}$).  This plot is for the choice ${m=3}$, ${a=1}$, ${l=0.2}$, ${e=g=1.6}$, and ${\alpha=0.12}$.
}
\label{Fig6}
\end{figure}

Although this condition is difficult to be solved analytically, some general observations can  be made. Clearly, the condition can not be satisfied in the regions where ${Q(r)<0}$. Naturally assuming a sufficiently small acceleration~$\alpha$ satisfying the inequality \eqref{orderofhorizonsexpl}, the function $P(\theta)$ is positive, while the  four distinct horizons are ordered as ${r_a^-<r_b^-<r_b^+<r_a^+}$, see \eqref{orderofhorizons}. For ${l<a}$, the metric function $Q$ satisfies  $Q(r)>0$ only in the regions $(r_a^-,r_b^-)$ and $(r_b^+,r_a^+)$, in which~$r$ is a \emph{spatial} coordinate. The closed timelike curves can thus \emph{only appear between the black hole horizon $\HH_b^\pm$ and the corresponding acceleration horizon $\HH_a^\pm$}, that is only in the region~IV given by ${r\in(r_a^-,r_b^-)}$ or in the region~II given by ${r\in(r_b^+,r_a^+)}$. On the contrary, the pathological domain can not occur in the region~III inside the black hole or close to the conformal infinities~$\scri^\pm$ which are the boundaries of the dynamical regions~I and~V where $r$ is temporal because ${Q<0}$. This fact is explicitly seen in the exact plots shown in Fig.~\ref{Fig6}.

Moreover, it can be proven analytically that \emph{these pathological regions with closed timelike curves do not overlap with the ergoregions} (shown in Fig.~\ref{Fig1}), although they are both in the same domains~II and~IV. Recall that the ergoregions are identified by the condition ${g_{tt}>0}$ (together with ${g_{rr}>0}$), that is
\begin{equation}
 Q < P \,a^2\sin^2\theta \,,
\label{gttergo}
\end{equation}
see Eq.~\eqref{gtt}. By substituting this inequality into  \eqref{gphiphi-general-condition}, which is the condition ${g_{\varphi\varphi} <0}$ for the pathological regions, we obtain the relation
\begin{equation}
 r^2+(a+l)^2 < a^2\sin^2\theta +4al\sin^2\!\tfrac{1}{2}\theta \,,
 \label{pathol-ergo}
\end{equation}
that is the same as ${\,r^2+a^2\cos^2\theta+2al\cos\theta+l^2 < 0\,}$. In view of \eqref{newrho}, we have thus obtained
\begin{equation}
\rho^2   \equiv r^2+(l+a \cos \theta)^2  <  0  \,,
 \label{pathol-ergo-fin}
\end{equation}
which is a contradiction.
\newpage

\vspace{0mm}
%

Interestingly, there is thus no intersection of the pathological regions with the ergoregions. This is in accord with a physical intuition: the pathological regions with closed timelike curves are located here in the vicinity of the twisting axis ${\theta=\pi}$, while the ergoregions are concentrated mostly near the equatorial plane ${\theta=\tfrac{\pi}{2}}$ of the rotating black hole horizons.

\subsection{Thermodynamic properties}
\label{subsec:thermodynamics}

Finally, we evaluate basic thermodynamic quantities of this class of black holes, namely the \emph{entropy}
\begin{equation}
S \equiv \frac{1}{4}\,{\cal A}\,,
\label{entropy}
\end{equation}
given by the horizon area ${\cal A}$, and the \emph{temperature}
\begin{equation}
T\equiv\frac{1}{2\pi}\,\kappa\,,
\label{temperature}
\end{equation}
given by the corresponding horizon surface gravity $\kappa$, see \cite{Wald:book1984}.

We obtain the \emph{horizon area} by integrating both angular coordinates of the metric \eqref{newmetricGP2005} for \emph{fixed values of $t$ and ${r=r_h}$},
\begin{equation}
\mathcal{A}(r_h) = \int_0^{2\pi C}\!\!\! \int_0^\pi \sqrt{g_{\theta \theta}\, g_{\varphi \varphi}}\,\,\dd \theta \, \dd \varphi\,,
\label{defA}
\end{equation}
where the metric functions are given by \eqref{gphiphi-gthetatheta0}. Using the fact that ${Q(r_h)=0}$ on any horizon, this expression simplifies to
\begin{equation}
\mathcal{A} = 2\pi C\,\big(r_h^2+(a+l)^2\big) \int_0^\pi \frac{\sin\theta}{\Omega^2(r_h)} \,\,\dd \theta \,.
\label{defAsimplified}
\end{equation}
Applying the explicit form of the conformal factor \eqref{newOmega}, an integration immediately leads to
\begin{align}\label{eq:HorizonArea}
\mathcal{A} = \frac{ 4\pi C\, \big(r_h^2+(a+l)^2\big)}
{\Big(1-\alpha\, {\displaystyle \frac{a^2+al}{a^2+l^2}}\,r_h  \Big)
 \Big(1+\alpha\, {\displaystyle \frac{a^2-al}{a^2+l^2}}\,r_h  \Big)}\,.
\end{align}
With the gauge \eqref{ChoiceOmega-al}, this is the same expression as Eq.~(51) in \cite{MatejovPodolsky:2021}.
In particular, for the \emph{four distinct horizons} $\HH$ introduced in \eqref{r+rep}--\eqref{ra-} we thus obtain that
\begin{eqnarray}
\hbox{area of} \ \HH_b^+ \ \hbox{is}&&  \mathcal{A}_b^+=\frac{ 4\pi C\, \big(r_{+}^2+(a+l)^2\big)}
{\Big(1-\alpha\, {\displaystyle \frac{a^2+al}{a^2+l^2}}\,r_{+}  \Big)\!
 \Big(1+\alpha\, {\displaystyle \frac{a^2-al}{a^2+l^2}}\,r_{+}  \Big)} \,, \label{Ar+}\\[1mm]
\hbox{area of} \ \HH_b^- \ \hbox{is}&&  \mathcal{A}_b^-=\frac{ 4\pi C\, \big(r_{-}^2+(a+l)^2\big)}
{\Big(1-\alpha\, {\displaystyle \frac{a^2+al}{a^2+l^2}}\,r_{-}  \Big)\!
 \Big(1+\alpha\, {\displaystyle \frac{a^2-al}{a^2+l^2}}\,r_{-}  \Big)} \,, \label{Ar-}\\[1mm]
\hbox{area of} \ \HH_a^+ \ \hbox{is}&&  \hbox{infinite} \,, \label{Ara+}\\[2mm]
\hbox{area of} \ \HH_a^- \ \hbox{is}&&  \hbox{infinite} \,. \label{Ara-}
\end{eqnarray}
The area of the acceleration horizons $\HH_a^\pm$ is thus \emph{unbounded}, while the black-hole horizons $\HH_b^\pm$ have \emph{finite values} given by \eqref{Ar+}, \eqref{Ar-}.

Interestingly, there exists a relation between these horizon areas and the conicities, namely
\begin{equation}
\mathcal{A}_b^+\mathcal{A}_b^- = 16\pi^2 C^2\,C_0\,C_\pi \, \big(r_{+}^2+(a+l)^2\big)\big(r_{-}^2+(a+l)^2\big) \,, \label{Ar+Ar-}
\end{equation}
where $C_0$ and $C_\pi$, given by \eqref{C0} and \eqref{Cpi}, are the specific conicities which regularize either the ${\theta=0}$ or the ${\theta=\pi}$ axis, respectively. For \emph{vanishing acceleration}~$\alpha$ the conicities are ${C=C_0=C_\pi=1}$, so that the two horizons of the complete family of Kerr--Newman--NUT black holes \eqref{metric-alpha=0}--\eqref{rho-alpha=0} located at
${r_{\pm} =  m \pm \sqrt{m^2 + l^2 - a^2 - e^2 - g^2}}$ have the corresponding areas
\begin{equation}
\mathcal{A}_b^\pm = 4\pi\, \big(r_{\pm}^2+(a+l)^2\big) \,. \label{Ar+-noacceleration}
\end{equation}
This simple expression reduces to the well-known formulas for Kerr--Newman black holes (${l=0}$), charged Taub--NUT (${a=0}$), Kerr (${l=0}$, ${e=0=g}$), Reissner--Nordstr\"{o}m (${a=0}$, ${l=0}$), and Schwarzschild (${a=0}$, ${l=0}$, ${e=0=g}$) with a single horizon of the area ${\mathcal{A}_b = 4\pi\, r_{h}^2 = 16\pi\,m^2 }$.

The \emph{surface gravity}~$\kappa$ is defined as the ``acceleration'' of the null normal $\xi^a$ generating the horizon at~$r_h$ via the relation ${\xi_{a;b}\,\xi^b  = \kappa\, \xi_a}$ (so that ${\kappa^2=-\frac{1}{2}\,\xi_{a;b}\,\xi^{a;b}}$). Previously in \cite{MatejovPodolsky:2021} we showed that for the general metric form \eqref{newmetricGP2005} this can be expressed as
\begin{equation}
\kappa = \frac{1}{2}\,\frac{Q'(r_h)}{r_h^2+(a+l)^2}\,,
\label{eq:kappa}
\end{equation}
where the prime denotes the derivative with respect to the coordinate~$r$. With the new factorized form \eqref{newQ} of the metric function $Q(r)$ this can now be easily evaluated, yielding
\begin{eqnarray}
\hbox{surface gravity of} \ \HH_b^+\, \hbox{is}\quad \kappa_b^+ \rovno \frac{\frac{1}{2}\big( r_{+}-r_{-} \big)
            \Big(1+\alpha\,{\displaystyle \frac{a^2-al}{a^2+l^2}}\, r_{+}\Big)\!
            \Big(1-\alpha\,{\displaystyle \frac{a^2+al}{a^2+l^2}}\, r_{+}\Big)}{r_+^2+(a+l)^2} , \label{kappa-b+}\\[1mm]
\hbox{surface gravity of} \ \HH_b^- \ \hbox{is}\quad  \kappa_b^- \rovno -\frac{\frac{1}{2}\big( r_{+}-r_{-} \big)
            \Big(1+\alpha\,{\displaystyle \frac{a^2-al}{a^2+l^2}}\, r_{-}\Big)\!
            \Big(1-\alpha\,{\displaystyle \frac{a^2+al}{a^2+l^2}}\, r_{-}\Big)}{r_-^2+(a+l)^2} , \label{kappa-b-}\\[1mm]
\hbox{surface gravity of} \ \HH_a^+ \ \hbox{is}\quad  \kappa_a^+ \rovno -\alpha\,\frac{a^2}{a^2+l^2}\,\frac{\big(r_a^+-r_{+} \big) \big( r_a^+-r_{-} \big)}{(r_a^+)^2+(a+l)^2} \,, \label{kappa-a+}\\[2mm]
\hbox{surface gravity of} \ \HH_a^- \ \hbox{is}\quad  \kappa_a^- \rovno \alpha\,\frac{a^2}{a^2+l^2}\,\frac{\big(r_a^--r_{+} \big) \big( r_a^--r_{-} \big)}{(r_a^-)^2+(a+l)^2} \,. \label{kappa-a-}
\end{eqnarray}
Recall that the specific values $r_{+}$, $r_{-}$, $r_a^+$, $r_a^-$ of the horizons position are explicitly given by \eqref{r+rep}--\eqref{ra-}. In particular,
\begin{equation}
\tfrac{1}{2}\big( r_{+}-r_{-} \big)  =  \sqrt{m^2 + l^2 - a^2 - e^2 - g^2}\,. \label{r+-r-}
\end{equation}

Notice that the surface gravities $\kappa$ (and thus the corresponding temperatures $T$) of the black-hole horizon $\HH_b^+$ and the acceleration horizon $\HH_a^-$ are \emph{positive}, while they are \emph{negative} for the complementary horizons $\HH_b^-$ and $\HH_a^+$.

It is also very interesting that even in the most general case the \emph{product of the area and the surface gravity of the black-hole horizons are the same}, and expressed simply as
\begin{equation}
 \mathcal{A}_b^+ \kappa_b^+ = -\mathcal{A}_b^- \kappa_b^- = 2\pi C \big( r_{+}-r_{-} \big)\,.
 \label{eq:Area*kappa}
\end{equation}
Consequently, the \emph{product of the temperature and the entropy} of the black-hole horizons $\HH_b^\pm$  is
\begin{equation}
 (TS)^+  = -(TS)^-  = \tfrac{1}{2} C\,\sqrt{m^2 + l^2 - a^2 - e^2 - g^2} \,.
 \label{eq:temp*entropy}
\end{equation}

Moreover, it is seen from \eqref{kappa-b+} and \eqref{kappa-b-} that
\begin{equation}
\kappa_b^+=0=\kappa_b^- \qquad \hbox{if and only if} \qquad r_+ = r_-
\label{eq:extremal-case}
\end{equation}
(assuming a reasonably small acceleration $\alpha$). This fully confirms that an \emph{extremal horizon has vanishing surface gravity}. As described in Sec.~\ref{sec_extreme}, if the extremality condition \eqref{extremality condition} is satisfied the double-degenerate extremal horizon is located at
\begin{equation}
r_h =  m\,, \label{r-extreme2}
\end{equation}
and the metric function $Q(r)$ takes the form \eqref{extremeQ},
\begin{eqnarray}
Q(r) \rovno (r-m)^2\,
            \Big(1+\alpha\,a\,\frac{a-l}{a^2+l^2}\, r\Big)
            \Big(1-\alpha\,a\,\frac{a+l}{a^2+l^2}\, r\Big). \label{extremeQ-repeated}
\end{eqnarray}
Clearly, ${Q(r_h) = 0}$ and also ${Q'(r_h) = 0}$, so that ${\kappa=0}$ due to \eqref{eq:kappa}.
Such a degenerate black-hole horizon at ${r=m}$ in the family of accelerating extremal  Kerr--Newman--NUT spacetimes has zero surface gravity, and thus zero thermodynamic temperature~$T$.

Let us consider the special case with \emph{vanishing acceleration} (${\alpha=0}$). In such a situation, the expressions \eqref{kappa-b+}--\eqref{kappa-a-} simplify:
\begin{eqnarray}
\hbox{surface gravity of} \ \HH_b^+\, \hbox{is}\quad \kappa_b^+ \rovno \frac{\sqrt{m^2 + l^2 - a^2 - e^2 - g^2}}{r_+^2+(a+l)^2} , \label{kappa-b+nonacc}\\[1mm]
\hbox{surface gravity of} \ \HH_b^- \ \hbox{is}\quad  \kappa_b^- \rovno -\frac{\sqrt{m^2 + l^2 - a^2 - e^2 - g^2}}{r_-^2+(a+l)^2} \,, \label{kappa-b-nonacc}\\[1mm]
\hbox{surface gravity of} \ \HH_a^\pm \ \hbox{is}\quad  \kappa_a^\pm \rovno 0 \,. \label{kappa-a-pm-nonacc}
\end{eqnarray}
(Actually, both the acceleration horizons $\HH_a^\pm$ disappear in this limit.) Writing \eqref{kappa-b+nonacc} fully explicitly, we obtain the surface gravity of the black-hole horizon $\HH_b^+$
\begin{eqnarray}
\kappa_b^+ \rovno \frac{\sqrt{m^2 + l^2 - a^2 - e^2 - g^2}}
{\big(\, m +\sqrt{m^2 + l^2 - a^2 - e^2 - g^2} \,\big)^2+(a+l)^2}\,. \label{kappa-b+nonacc-explicit}
\end{eqnarray}
This generalizes for the case ${l\ne0}$ and ${g\ne0}$ the expression
\begin{equation}
\kappa = \frac{\sqrt{m^2 - a^2 - e^2}}{2m(m+\sqrt{m^2 - a^2 - e^2})-e^2}\,,
\label{surface_gravity_KN}
\end{equation}
which is the usual surface-gravity formula for the Kerr--Newman black hole, see Eq.~(12.5.4) in \cite{Wald:book1984}. For the Schwarzschild black hole it reads ${\kappa = 1/(4m)}$.

Finally, let us remark that our explicit and fully general expressions \eqref{kappa-b+}--\eqref{kappa-a-} for the surface gravity~$\kappa$ of each of the 4 horizons at~$r_h$ agree with the results obtained \emph{directly} from the definition ${\xi_{a;b}\,\xi^b  = \kappa\, \xi_a}$ if the appropriate null normal generator $\xi^a$ of the horizon is employed. In particular, the corresponding Killing vector field is
\begin{eqnarray}
\xi^a \equiv \partial_t + \Omega_h \, \partial_\varphi\,,
\end{eqnarray}
where the constant $\Omega_h$ is the \emph{angular velocity of the given horizon~$\HH$}.
Using \eqref{gtt-gtphi0} and \eqref{gphiphi-gthetatheta0}, the norm $\xi^a\xi_a$ of the Killing vector $\xi^a$ at the horizon (where ${Q=0}$)  vanishes if and only if
\begin{eqnarray}
\Omega_h \rovno \frac{a}{r_h ^2 +(a+l)^2} \,. \label{Omega-h}
\end{eqnarray}
For the particular horizons ${r_b^\pm \equiv r_{\pm}}$ and ${r_a^\pm}$ given by \eqref{r+rep}--\eqref{ra-} this gives the constants
\begin{eqnarray}
\Omega^\pm_b \rovno \frac{a}{r_\pm ^2 +(a+l)^2} \,, \label{Omega-b}\\
\Omega^\pm_a \rovno \frac{\alpha^2 a^3 \, (a \pm l)^2 }{(a^2+l^2)^2+ {\alpha}^2 a^2 \, (a+l)^2 (a \pm l)^2 } \label{Omega-a}\,.
\end{eqnarray}

It can be seen that for vanishing Kerr-like rotation (${a=0}$) the angular velocities of all four horizons become zero, whereas for vanishing NUT parameter (${l=0}$) they all remain non-zero,
\begin{equation}
\Omega^\pm_b = \frac{a}{r_\pm ^2 +a^2} \,, \qquad
\Omega^\pm_a = \frac{\alpha^2 a}{1+ \alpha^2 a^2 } \label{Omega-l=0veloc}\,.
\end{equation}

\newpage

\subsection{Concluding summary}
\label{sec:summary}

In this work we presented a new metric form \eqref{newmetricGP2005}--\eqref{r-} of the remarkable family of exact black holes of algebraic type~D, initially found by Debever (1971) and by Pleba\'nski and Demia\'nski (1976). Moreover, we demonstrated that this improved metric representation has many advantages which simplify the investigation of its geometrical and physical properties. In particular:

\begin{itemize}

\item In Sec.~\ref{sec_derivation} we started with a convenient Griffiths--Podolsk\'y (2005, 2006) form of this class of spacetimes, but we further improved it. By introducing a modified set of the mass and charge parameters ${m, e, g}$, applying a special conformal rescaling $S$, and choosing a useful gauge of the twist parameter $\omega$, we obtained an explicit compact form of the metric.

\item The metric functions \eqref{newOmega}--\eqref{newQ} are very simple, depending only on the radial coordinate $r$ and the angular coordinate $\theta$. Moreover, the key functions $P(\theta)$ and $Q(r)$ are factorized. They explicitly localize the axes of symmetry and the horizons, respectively.

\item The metric depends on six parameters $m, a, l, \alpha, e, g$ with direct physical meaning, namely they represent the mass, Kerr-like rotation, NUT parameter, acceleration, electric, and magnetic charges of the black hole, respectively.

\item Interestingly, the new metric \eqref{newmetricGP2005} depends on the parameters $a, l, \alpha$ directly, while the dependence on the remaining three parameters $m, e, g$ is encoded in the two constants $r_{+}$ and $r_{-}$ defined by \eqref{r+} and \eqref{r-}. In fact, these expressions localize the two black-hole horizons, and they only appear in the factorized metric functions $P$ and $Q$.

\item Very nice feature of the new metric form \eqref{newmetricGP2005}--\eqref{newQ} is that any of its six physical parameters can be independently set to zero, and this can be done in any order. In this way, specific subclasses of type~D black holes are easily obtained.

\item This property is demonstrated in Sec.~\ref{sec_subclasse} where the general family of accelerating, charged, rotating and NUTed black holes naturally reduce to its large subclasses with five physical parameters. These are the Kerr--Newman--NUT black holes without acceleration (${\alpha=0}$), accelerating Kerr--Newman black holes without NUT (${l=0}$), charged Taub--NUT black holes without rotation (${a=0}$), and accelerating Kerr--NUT black holes without electric or magnetic charges (${e=0}$ or ${g=0}$).

\item All the metric functions \eqref{newOmega}--\eqref{newQ} depend on the acceleration $\alpha$ only via the product ${\alpha\,a}$. Therefore, by setting the Kerr-like rotation $a$ to zero, the new metric \eqref{newmetricGP2005} becomes independent of~$\alpha$, and simplifies directly to charged Taub--NUT black holes. This explicitly confirms the previous observation made by Griffiths and Podolsk\'y that there is no accelerating NUT black hole in the Pleba\'nski--Demia\'nski family of type~D spacetimes. Quite surprisingly, such a solution for accelerating non-rotating black hole with purely NUT parameter exists\cite{ChngMannStelea:2006,PodolskyVratny:2020}, but it is of distinct algebraic type~I.

\item The simplest subcases of our general metric \eqref{newmetricGP2005} with just the mass $m$ and one additional physical parameter reveal the famous black holes, namely the Schwarzschild, Reissner--Nordstr\"{o}m, Kerr, Taub--NUT or the $C$-metric solutions, all in their standard coordinate forms.

\item As shown in Sec.~\ref{sec_extreme}, the improved metric \eqref{newmetricGP2005} naturally contains also extreme black holes with double-degenerate horizons (${r_{+}=r_{-}}$) located at ${r = m}$, whenever ${m^2 + l^2 =  a^2 + e^2 + g^2 }$. Such a family of extremal accelerating Kerr--Newman--NUT black holes also admits various subclasses, obtained by setting any of the  parameters $\alpha, l, a, e, g$ to zero. In fact, they represent the complete class of extremal isolated horizons with axial symmetry \cite{MatejovPodolsky:2021}.

\item The hyperextreme cases, when the parameters satisfy the relation ${m^2 + l^2 <  a^2 + e^2 + g^2}$, represent exact spacetimes with an accelerated naked singularity. The metric functions $P, Q$ are not (fully) factorizable, and take the form \eqref{hyperextremeP}, \eqref{hyperextremeQ}. There are thus only two acceleration horizons, which are absent when ${\alpha\,a=0}$.

\end{itemize}

The new convenient metric \eqref{newmetricGP2005} considerably simplifies the investigation of various properties of this large family of black holes, as demonstrated in the subsequent sections of our work, namely:

\begin{itemize}

\item First, in Sec.~\ref{sec_discussion} we evaluated the Weyl and Ricci tensors of \eqref{newmetricGP2005}, expressed as the Newman--Penrose scalars in the natural tetrad
    \eqref{nullframe} adapted to the double-degenerate principal null directions. The only such scalars are
    $\Psi_2$ and $\Phi_{11}$, confirming the type~D algebraic structure of the gravitational field, aligned with the non-null electromagnetic field \eqref{vector-potential}--\eqref{Phi1}.

\item  Their explicit form \eqref{Psi2} and \eqref{Phi11} reveals that generic black-hole spacetimes are asymptotically flat at ${\Omega = 0}$. For vanishing acceleration~$\alpha$, the spacetimes \eqref{newmetricGP2005} become asymptotically flat for large values of the radial coordinate~$|r|$ (except along the axes of symmetry
    ${\theta=0}$ and ${\theta=\pi}$ if the cosmic strings or struts are present).

\item Both the double-degenerate principal null directions are expanding. They are twisting if and only if ${a=0=l}$. On the horizons, the expansion and twist always vanish.

\item In general, there are four distinct horizons identified in Subsec.~\ref{subsec:horizon} as the roots of the metric function $Q(r)$. Since its form \eqref{newQrep} is fully factorized, the corresponding positions are simply expressed in terms of the physical parameters as \eqref{r+rep}--\eqref{ra-}. There is a pair of black-hole horizons $\HH_b^\pm$ at ${r_b^\pm \equiv r_\pm = m \pm \sqrt{m^2 + l^2 - a^2 - e^2 - g^2}}$, and a pair of acceleration horizons $\HH_a^\pm$ at ${r_a^\pm \equiv \pm \,\alpha^{-1}(a^2+l^2)/(a^2\pm a\,l)}$, which simplifies to ${r_a^\pm \equiv \pm \,\alpha^{-1}}$ when ${l=0}$.

\item Interestingly, these positions of the black-hole horizons are independent of the acceleration~$\alpha$, while the acceleration horizons do not depend on the mass $m$ and the charges $e,g$.

\item For sufficiently small acceleration $\alpha$ such that ${\alpha\,r_+ < (a^2+l^2)/(a^2+a\,l)}$, with ${0\le l <a}$, the four horizons are ordered as ${r_a^-<r_b^-<r_b^+<r_a^+}$, see \eqref{orderofhorizonsexpl}.

\item Whenever the Kerr-like rotation parameter~$a$ is non-zero, each of these four horizons is accompanied by the corresponding ergoregion, see Subsec.~\ref{subsec:ergoregions}. It ``touches' the horizon at its poles, extending from the horizon near the equatorial region. This is shown in Fig.~\ref{Fig1}. For the Kerr--Newman--NUT black holes without acceleration, the ergoregions are bounded by the surface ${r_{e\pm}(\theta) = m \pm \sqrt{m^2 + l^2 - e^2 - g^2 - a^2\cos^2\theta }}$.

\item Using the Weyl scalar $\Psi_2$ and also the Kretschmann scalar ${\mathcal{K} \equiv R_{abcd}\, R^{abcd}}$, in Subsec.~\ref{subsec:singularities} we clarified the presence and the structure of the curvature singularities. Such a singularity is present at ${r = 0}$, but only if ${l+a \cos \theta = 0}$ which requires ${|l| \leq |a|}$. There is thus no curvature singularity in the black-hole spacetimes with large NUT parameter ${|l|>|a|\ge 0}$.

\item For ${0\le|l|\le|a|}$ the curvature singularity is present at ${r=0}$, but only in the section with special value of the angular coordinate $\theta$ such that ${\cos \theta = -l/a}$. Various possibilities are summarized in \eqref{a-versus-a}.

\item This singularity has a ring structure which can be crossed from the asymptotically flat region ${r>0}$ to the distinct asymptotically flat region ${r<0}$, as schematically shown in Fig.~\ref{Fig2} and Fig.~\ref{Fig3}. Only in the section ${\cos \theta = -l/a}$ (or for any value of~$\theta$ if ${l=0=a}$) we have to restrict the range of~$r$ to two separate domains ${r>0}$ and ${r<0}$.

\item To complete our understanding of the global causal structure of the entire family of black-hole spacetimes \eqref{newmetricGP2005}, in Subsec.~\ref{subsec:global} we introduced the retarded and advanced null coordinates in which the corresponding metric forms \eqref{metric-advanced} and \eqref{metric-retarded} have no coordinate singularities at the horizons.

\item Then we explicitly constructed the corresponding Kruskal--Szekeres-type coordinates which enabled us to perform  the maximal analytic extension across all the horizons. It revealed an infinite number of time-dependent regions (of type I, III, V) and stationary regions (of type II, IV) which are separated by the black hole and acceleration horizons $\HH_b^\pm$ and $\HH_a^\pm$.

\item The complicated global structure of this large family of spacetimes is visualized in the Penrose diagrams obtained by a suitable conformal compactification, drawn in Fig.~\ref{Fig4} and Fig.~\ref{Fig5}. The complete manifold contains an infinite number of black holes in various asymptotically flat universes identified by  distinct (future and past) conformal infinities $\scri$ --- unless a special topological identification is made.

\item In Subsec.~\ref{subsec:strings} we clarified that the physical source of acceleration of the black holes is the tension (or compression) in the rotating cosmic strings (or struts) located along the two axes of axial symmetry at ${\theta=0}$ and ${\theta=\pi}$. Such strings or struts are related to the deficit or excess angles which introduce topological defects along these axes (while the curvature remains finite).

\item In general, there are strings/struts along both the axes, but one of the axis can be made fully regular by a suitable choice of the conicity parameter $C$ in the range ${\varphi\in[0,2\pi C)}$. The first axis ${\theta=0}$ is regular in the metric form \eqref{newmetricGP2005} with the choice \eqref{C0}, whereas the second axis ${\theta=\pi}$ is regular in the form \eqref{newmetricGP2005-Regular-pi} with the choice \eqref{Cpi}. In the first case, there is a cosmic strut along ${\theta=\pi}$ with the excess angle \eqref{deltapi}, while in the second case there is a cosmic string along ${\theta=0}$ with the deficit angle \eqref{delta0}. For vanishing acceleration, both the axes can be made regular simultaneously (except for a possible NUT-like pathology).

\item In addition to the deficit/excess angles, these cosmic strings/struts located along the axes of symmetry are characterized by their rotation parameter $\omega$ (angular velocity). We demonstrated in Subsec.~\ref{subsec:rotatingstrings} that their values are directly related to the NUT parameter $l$, see expressions (\ref{omega0pi}) and (\ref{omega0piregpi}).

\item There is always a constant difference ${\Delta\omega=4l}$ between the angular velocities of the two rotating cosmic strings or struts. If and only if ${l=0}$, both the axes are nontwisting.

\item In the neighborhood of these rotating strings/struts there occur pathological regions with closed timelike curves. As shown in Subsec.~\ref{subsec:CTCrotatingstrings}, these regions are generally given by the condition  \eqref{gphiphi-general-condition}. They appear close to the rotating strings/struts, but only between the black hole horizon $\HH_b^\pm$ and the corresponding acceleration horizon $\HH_a^\pm$ (that is in the domains of type~II and~IV), see Fig.~\ref{Fig6}.

\item Although the pathological regions with closed timelike curves are located in the same domains as the ergoregions, they do not overlap with each other.

\item The convenient metric form \eqref{newmetricGP2005} with straightforward identification of the horizons is also suitable for an easy investigation of the black hole thermodynamics. Indeed, in Subsec.~\ref{subsec:thermodynamics} we explicitly evaluated the area of the four horizons \eqref{Ar+}--\eqref{Ara-}, their surface gravity \eqref{kappa-b+}--\eqref{kappa-a-}, and their angular velocity \eqref{Omega-b}--\eqref{Omega-a}.

\item These expressions generalize the usual formulas for the Kerr--Newman family to black holes with acceleration $\alpha$ and NUT parameter~$l$. They reveal interesting relations for the horizons temperature and entropy, for example ${(TS)^+  = -(TS)^-  = \tfrac{1}{2} C\,\sqrt{m^2 + l^2 - a^2 - e^2 - g^2}}$.

\end{itemize}

To conclude, the simple new metric form \eqref{newmetricGP2005}--\eqref{r-} has clear advantages. We hope that it will be employed for various studies and applications of this interesting class of accelerating and rotating black holes which charges and the NUT parameter.

\section*{Acknowledgments}

This work has been supported by the Czech Science Foundation Grant No.~GA\v{C}R 20-05421S (JP) and by the Charles University project GAUK No.~358921 (AV).

\end{document}